\documentclass[preprint]{ptephy}

\preprintnumber{KYUSHU-HET-175}

\usepackage{amssymb}
\usepackage{amsthm}
\usepackage{amsmath}
\usepackage{booktabs}
\usepackage{bbm}
\usepackage{mathtools}
\DeclareMathOperator{\tr}{tr}

\newcommand{\Slash}[1]{{\ooalign{\hfil/\hfil\crcr$#1$}}}
\numberwithin{equation}{section}

\usepackage{hyperref}

\begin{document}

\title{4D $\mathcal{N}=1$ SYM supercurrent in terms of the gradient flow}

\author{%
\name{\fname{Kenji} \surname{Hieda}}{1},
\name{\fname{Aya} \surname{Kasai}}{1},
\name{\fname{Hiroki} \surname{Makino}}{1},
and
\name{\fname{Hiroshi} \surname{Suzuki}}{1,\ast}
}

\address{%
\affil{1}{Department of Physics, Kyushu University
744 Motooka, Nishi-ku, Fukuoka, 819-0395, Japan}
\email{hsuzuki@phys.kyushu-u.ac.jp}
}

\begin{abstract}
The gradient flow and its small flow-time expansion provide a very versatile
method to represent renormalized composite operators in a
regularization-independent manner. This technique has been utilized to
construct typical Noether currents such as the energy--momentum tensor and the
axial-vector current in lattice gauge theory. In this paper, we apply the same
technique to the supercurrent in the four-dimensional $\mathcal{N}=1$ super
Yang--Mills theory (4D $\mathcal{N}=1$ SYM) in the Wess--Zumino gauge. Since
this approach provides a priori a representation of the properly normalized
conserved supercurrent, our result should be useful, e.g., in lattice
numerical simulations of the 4D $\mathcal{N}=1$ SYM; the conservation of the
so-constructed supercurrent can be used as a criterion for the supersymmetric
point toward which the gluino mass is tuned.
\end{abstract}

\subjectindex{B01, B16, B31, B32}
\maketitle

\section{Introduction}
\label{sec:1}
The so-called gradient flow (Refs.~\cite{Narayanan:2006rf,Luscher:2009eq,%
Luscher:2010iy,Luscher:2011bx,Luscher:2013cpa}) possesses a remarkable
renormalization property that any local product (i.e., composite operator)
composed from \emph{bare fields\/} evolved by the flow automatically becomes a
renormalized composite operator (Refs.~\cite{Luscher:2011bx,Luscher:2013cpa,%
Hieda:2016xpq}).\footnote{Up to the wave-function renormalization of the flowed
matter fields; see below.} By utilizing this property, one can express physical
quantities, such as a nonperturbative gauge coupling
(Refs.~\cite{Luscher:2011bx,Borsanyi:2012zs,Fodor:2012td,Fritzsch:2013je}),
the topological charge (Refs.~\cite{Luscher:2010iy,Ce:2015qha}), the chiral
condensate (Ref.~\cite{Luscher:2013cpa}), and the quasi-parton distribution
functions (Ref.~\cite{Monahan:2016bvm} etc.\ in terms of bare fields evolved by
the flow. Such finite representations of physical quantities are universal,
i.e., independent of the regularization and are particularly useful in the
context of lattice gauge theory. See Ref.~\cite{Luscher:2013vga} for a review.
In general, however, to find such a representation of a desired physical
quantity in terms of the flowed fields requires an ingenious argument quantity
by quantity.

On the other hand, one can always employ the small flow-time expansion
(Ref.~\cite{Luscher:2011bx}) to express any renormalized local composite
operator in terms of the flowed fields. The combination of the gradient flow
and the small flow-time expansion thus provides a very versatile method to
represent a renormalized composite operator in a regularization-independent
manner. This technique has been utilized to construct typical Noether currents
such as the energy--momentum tensor
(Refs.~\cite{Suzuki:2013gza,Makino:2014taa,%
Suzuki:2016ytc}) (see also Ref.~\cite{DelDebbio:2013zaa} for a related study)
and the axial-vector current (Refs.~\cite{Endo:2015iea,Hieda:2016lly}).%
\footnote{In~Refs.~\cite{Endo:2015iea,Hieda:2016lly}, the vector current and
the (pseudo-) scalar density are also studied.} The resulting representations
have been numerically examined/applied in the quenched and $2+1$~flavor
QCD (Refs.~\cite{Asakawa:2013laa,Taniguchi:2016ofw,Kitazawa:2016dsl,%
Taniguchi:2016tjc}) and analytically examined in some solvable
models (Refs.~\cite{Makino:2014sta,Makino:2014cxa,Suzuki:2015fka}).

In this paper, we apply the above technique to find a representation of the
supercurrent in the four-dimensional $\mathcal{N}=1$ super Yang--Mills theory
(4D $\mathcal{N}=1$ SYM) (Refs.~\cite{Wess:1974jb,Ferrara:1974pu,Salam:1974jj,%
Wess:1992cp}). A generalization of the gradient flow to this system in its
off-shell supermultiplet was developed in~Ref.~\cite{Kikuchi:2014rla} by
partially aiming at identical renormalizations between the flowed gauge field
and the flowed fermion (i.e., gaugino) field through the off-shell
supersymmetry. Our objective here is much modest: Having the application in the
conventional lattice gauge theory in mind, we take the Wess--Zumino gauge,
which contains only conventional dynamical fields but preserves only the
on-shell supersymmetry. Since our approach provides a priori the
properly normalized and conserved supercurrent, our result should be useful,
e.g., in lattice numerical simulations of the 4D $\mathcal{N}=1$
SYM (Refs.~\cite{Curci:1986sm,Kaplan:1983sk,Kaplan:1992bt,Montvay:1995ea,%
Nishimura:1997vg,Maru:1997kh,Donini:1997hh,Neuberger:1997bg,Kirchner:1998mp,%
Campos:1999du,Taniguchi:1999fc,Kaplan:1999jn,Fleming:2000fa,Farchioni:2001wx,%
Montvay:2001aj,Farchioni:2004fy,Giedt:2008xm,Endres:2009yp,Demmouche:2010sf,%
Kim:2011fw,Suzuki:2012pc,Bergner:2015adz,Ali:2016zke})
(Ref.~\cite{Kadoh:2016eju} is a recent review); the conservation of the
so-constructed supercurrent can be used as a criterion for the supersymmetric
point toward which the gluino mass is tuned. For the 4D $\mathcal{N}=1$ SYM,
one may alternatively invoke the chiral symmetry to find the supersymmetric
point. A tuning of parameters to the supersymmetric point will really become
demanding, however, in supersymmetric theories containing matter multiplets. In
such theories, a priori knowledge of the conserved supercurrent will be quite
helpful to find the supersymmetric point. Thus, the present work can be
regarded as the first step toward a flow-time representation of the
supercurrent in such complicated supersymmetric systems.

Now, to find a representation of the properly normalized conserved supercurrent
in terms of the flowed fields, we have to know the expression of the former at
least in perturbation theory. For the energy--momentum tensor
(Refs.~\cite{Suzuki:2013gza,Makino:2014taa}), the all-order expression is
readily available in dimensional regularization (see, e.g.,
Refs.~\cite{Nielsen:1977sy,Collins:1976yq}) because this
regularization preserves the translational invariance exactly. For the
axial-vector current, the naive expression in dimensional regularization must
be corrected appropriately (Ref.~\cite{Collins:1984xc}) so that it fulfills the
corresponding Ward--Takahashi (WT) relations. It is not difficult to carry out
this correction, at least in the one-loop order, because the chiral
transformation acts only on the fermion fields only linearly;
in~Refs.~\cite{Endo:2015iea,Hieda:2016lly}, the one-loop corrected expression
in the dimensional regularization was employed.

Unfortunately, the situation is much more complicated for the supercurrent in
the Wess--Zumino gauge. First, there is no regularization that manifestly
preserves supersymmetry; we thus adopt dimensional regularization in what
follows for computational convenience.\footnote{One might think that the use
of the dimensional reduction (Ref.~\cite{Siegel:1979wq}) rather simplifies our
task. However, since the Fierz identity must be given up with the dimensional
reduction (Refs.~\cite{Avdeev:1981vf,Avdeev:1982xy}), we could not find that
the dimensional reduction is particularly useful for our purpose.} Second, the
super transformation acts on both the gauge field and the fermion field. Then
the WT relations necessarily contain contributions from the gauge-fixing term
and the Faddeev--Popov ghost term, which are neither gauge invariant nor
supersymmetric in the Wess--Zumino gauge. Finally, the super transformation is
nonlinear in the Wess--Zumino gauge and thus the WT relations necessarily
contain composite operators. Even though the gauge field and the fermion field
are related by supersymmetry, the wave-function renormalization factors for
those fields \emph{differ\/} in the Wess--Zumino gauge. The validity of
supersymmetry WT relations in the Wess-Zumino gauge thus crucially depends on
the renormalization of composite operators appearing in the WT relations. The
fact that gauge invariance is broken by the gauge-fixing term and the
Faddeev--Popov term further complicate the situation, because one has also to
take into account the operator mixing with gauge noninvariant operators.
In short, to find the correct expression of the supercurrent, one has to fully
understand the renormalization/mixing structure of composite operators
appearing in supersymmetry WT relations.

Somewhat surprisingly, to our knowledge, this program to find an explicit form
of the supercurrent in the 4D $\mathcal{N}=1$ SYM in the Wess--Zumino gauge
(e.g., in dimensional regularization) has not been carried out thoroughly in
the literature. The only exception we could find is Ref.~\cite{Howe:1984zt} but
in this reference only the case of the abelian gauge theory is studied. Thus,
we had to carry out this program by ourselves; Sect.~\ref{sec:2} is devoted to
this complicated analysis in the one-loop order. Our conclusion is that in
dimensional regularization, the properly normalized supercurrent in correlation
functions of gauge invariant operators (which do not contain the ghost field
and the anti-ghost field) is given by\footnote{Notation: Without noting
otherwise, repeated indices are understood to be summed over. The
generators~$T^a$ of the gauge group~$G$ are anti-Hermitian and the structure
constants are defined by~$[T^a,T^b]=f^{abc}T^c$. Quadratic Casimirs are defined
by~$f^{acd}f^{bcd}=C_2(G)\delta^{ab}$ and, for a gauge representation~$R$,
$\tr_R(T^aT^b)=-T(R)\delta^{ab}$ and~$T^aT^a=-C_2(R)1$. We also denote
$\tr_R(1)=\dim(R)$. For the fundamental $N$~representation of~$SU(N)$ for which
$\dim(N)=N$, our choice is
\begin{equation}
   C_2(SU(N))=N,\qquad T(N)=\frac{1}{2},\qquad
   C_2(N)=\frac{N^2-1}{2N}.
\label{eq:(1.1)}
\end{equation}
Our Dirac matrices~$\gamma_\mu$ are all Hermitian and for the trace over the
spinor index we set $\tr(1)=4$ for any spacetime dimension~$D$. The chiral
matrix is defined by~$\gamma_5=\gamma_0\gamma_1\gamma_2\gamma_3$ for any~$D$.
We also use the symbol $\sigma_{\mu\nu}\equiv(1/2)[\gamma_\mu,\gamma_\nu]$.}
\begin{equation}
   S_{\mu R}(x)
   =-\frac{1}{2g_0}\sigma_{\rho\sigma}\gamma_\mu
   \psi^a(x)F_{\rho\sigma}^a(x)+\mathcal{O}(g_0^3),
\label{eq:(1.2)}
\end{equation}
where $g_0$ is the bare gauge coupling, $\psi^a(x)$ is the bare gaugino field
and
$F_{\rho\sigma}^a(x)\equiv\partial_\rho A_\sigma^a(x)-\partial_\sigma A_\rho^a(x)
+f^{abc}A_\rho^b(x)A_\sigma^c(x)$ is the bare field strength. The expression to
this order is thus identical to the naive classical expression of the
supercurrent; we emphasize however that Eq.~\eqref{eq:(1.2)} is the result of a
lengthy analysis of the renormalization/mixing of composite operators in
supersymmetry WT relations.

Once Eq.~\eqref{eq:(1.2)} is obtained, it is basically straightforward to
express the supercurrent in terms of the small flow-time limit of flowed
fields. This is done in~Sect.~\ref{sec:3}; after some calculation, we
have\footnote{Although $\Bar{g}(1/\sqrt{8t})\to0$ as~$t\to0$ due to the
asymptotic freedom, $\mathcal{O}(\Bar{g}(1/\sqrt{8t}))$ terms still provide
useful information because they tell us how the representation approaches
the real supercurrent as~$t\to0$; we cannot simply set $t\to0$ in lattice
numerical simulations with finite lattice spacings.
See~Refs.~\cite{Asakawa:2013laa,Taniguchi:2016ofw,Kitazawa:2016dsl} for the
situation for a similar representation of the energy--momentum tensor.}
\begin{align}
   S_{\mu R}(x)
   &=\lim_{t\to0}\biggl(
   -\frac{1}{2\Bar{g}(1/\sqrt{8t})}
   \left\{1+\frac{\Bar{g}(1/\sqrt{8t})^2}{(4\pi)^2}C_2(G)
   \left[-\frac{7}{2}-\frac{3}{2}\ln\pi+\frac{1}{2}\ln(432)\right]
   \right\}
\notag\\
   &\qquad\qquad\qquad\qquad\qquad\qquad\qquad\qquad\qquad\qquad\qquad{}
   \times\sigma_{\rho\sigma}\gamma_\mu\mathring{\chi}^a(t,x)G_{\rho\sigma}^a(t,x)
\notag\\
   &\qquad\qquad{}
   -\frac{\Bar{g}(1/\sqrt{8t})}{(4\pi)^2}C_2(G)3
   \gamma_\nu\mathring{\chi}^a(t,x)G_{\nu\mu}^a(t,x)\biggr).
\label{eq:(1.3)}
\end{align}
In this expression, $\Bar{g}(\mu)$, $\mathring{\chi}^a(t,x)$,
and~$G_{\rho\sigma}^a(t,x)$ are the running gauge coupling in the minimal
subtraction (MS) scheme at the renormalization scale~$\mu$, the flowed gaugino
field, and the flowed field strength, respectively; the precise definitions of
these quantities will be given in~Sect.~\ref{sec:3}. At this point, it is
interesting to note that the above expression reproduces the gamma-trace
anomaly (superconformal anomaly)~\cite{Abbott:1977in,Curtright:1977cg,%
Inagaki:1978iu,Majumdar:1980ej,Nicolai:1980km,Hagiwara:1979pu,Hagiwara:1980ys,%
Kumar:1982ng,Nakayama:1983qt}
$\gamma_\mu S_{\mu R}(x)=-g/(4\pi)^2C_2(G)3\sigma_{\mu\nu}\psi^a(x)F_{\nu\mu}^a(x)
+\mathcal{O}(g^3)$ at least in the one-loop order, because
$\gamma_\mu\sigma_{\rho\sigma}\gamma_\mu=0$ for~$D=4$ and flowed fields simply
reduce to the corresponding un-flowed fields in the $t\to0$ limit in the lowest
order of perturbation theory. This is reassuring, because the properly
normalized conserved supercurrent must possess this gamma-trace
anomaly.\footnote{In Appendix~\ref{app:B}, by using the information of the
superconformal anomaly to the two-loop order, we further improve the small
flow-time representation~\eqref{eq:(1.3)}.}

Section~\ref{sec:4} is devoted to the conclusion. In~Appendix~\ref{app:A}, we
clarify how the charge conjugation matrix should be treated in dimensional
regularization, because a description of this issue is also hard to find in
the literature.

\section{Properly normalized conserved supercurrent in dimensional
regularization}
\label{sec:2}
\subsection{Actions, the super transformation and the BRS transformation}
In what follows, we denote the spacetime dimension as~$D\equiv4-2\epsilon$,
assuming dimensional regularization. The classical action of the
$\mathcal{N}=1$ SYM in the Wess--Zumino gauge is given by
\begin{equation}
   S=\frac{1}{4g_0^2}\int d^Dx\,F_{\mu\nu}^a(x)F_{\mu\nu}^a(x)
   +\frac{1}{2}\int d^Dx\,\Bar{\psi}^a(x)\Slash{\mathcal{D}}^{ab}\psi^b(x),
\label{eq:(2.1)}
\end{equation}
where the covariant derivative is defined in the adjoint representation,
\begin{align}
   \mathcal{D}_\mu^{ab}
   &\equiv\delta^{ab}\partial_\mu+f^{acb}A_\mu^c(x)
\notag\\
   &\equiv\delta^{ab}\partial_\mu+\mathcal{A}_\mu^{ab}(x),
\label{eq:(2.2)}
\end{align}
and $\Slash{\mathcal{D}}^{ab}\equiv\gamma_\mu\mathcal{D}_\mu^{ab}$.

The gaugino field~$\psi(x)$ is the Majorana spinor. This implies that
$\Bar{\psi}(x)$ is not an independent dynamical variable and it is given
from~$\psi(x)$ by
\begin{equation}
   \Bar{\psi}(x)=\psi^T(x)(-C^{-1}),
\label{eq:(2.3)}
\end{equation}
where $T$ denotes the transpose on the Dirac index and $C$ is the charge
conjugation matrix. The properties of~$C$ in dimensional regularization are
summarized in~Appendix~\ref{app:A}. In particular, the matrix $C^{-1}\gamma_\mu$
is symmetric in its Dirac indices for any~$D$; this is crucial for the
action~\eqref{eq:(2.1)} to be meaningful for any~$D$.

The super transformation in the Wess-Zumino gauge is given by
\begin{align}
   \delta_\xi A_\mu(x)&=g_0\Bar{\xi}\gamma_\mu\psi(x),&&
\label{eq:(2.4)}
\\
   \delta_\xi\psi(x)&=-\frac{1}{2g_0}\sigma_{\mu\nu}\xi F_{\mu\nu}(x),&
   \delta_\xi\Bar{\psi}(x)&=\frac{1}{2g_0}\Bar{\xi}\sigma_{\mu\nu}F_{\mu\nu}(x),
\label{eq:(2.5)}
\end{align}
where the parameter~$\xi$ is also the Majorana spinor:
\begin{equation}
   \Bar{\xi}=\xi^T(-C^{-1}).
\label{eq:(2.6)}
\end{equation}
The classical action~\eqref{eq:(2.1)} with~$D=4$ is invariant under this
transformation (see below).

To carry out perturbation theory, we also introduce the gauge-fixing term
\begin{equation}
   S_{\text{gf}}=\frac{\lambda_0}{2g_0^2}
   \int d^Dx\,\partial_\mu A_\mu^a(x)\partial_\nu A_\nu^a(x),
\label{eq:(2.7)}
\end{equation}
where $\lambda_0$ is the bare gauge-fixing parameter and the corresponding
Faddeev--Popov ghost term
\begin{equation}
   S_{c\Bar{c}}=-\frac{1}{g_0^2}\int d^Dx\,\Bar{c}^a(x)\partial_\mu
   \mathcal{D}_\mu^{ab}c^b(x).
\label{eq:(2.8)}
\end{equation}
$c^b(x)$ is the ghost field and $\Bar{c}^a(x)$ is the anti-ghost field. Then
the whole action~$S+S_{\text{gf}}+S_{c\Bar{c}}$ is invariant under the following
BRS transformation:
\begin{align}
   \Hat{\delta}_B A_\mu^a(x)&=\mathcal{D}_\mu^{ab}c^b(x),&
   \Hat{\delta}_B c^a(x)&=-\frac{1}{2}f^{abc}c^b(x)c^c(x),
\label{eq:(2.9)}
\\
   \Hat{\delta}_B\Bar{c}^a(x)&=\lambda_0\partial_\mu A_\mu^a(x),&&
\label{eq:(2.10)}
\\
   \Hat{\delta}_B\psi^a(x)&=-f^{abc}c^b(x)\psi^c(x),&
   \Hat{\delta}_B\Bar{\psi}^a(x)&=-f^{abc}c^b(x)\Bar{\psi}^c(x).
\label{eq:(2.11)}
\end{align}

\subsection{Bare Ward--Takahashi relations in the gauge-fixed theory}
Now, the classical expression of the supercurrent can be found by making the
parameter~$\xi$ in~Eqs.~\eqref{eq:(2.4)} and~\eqref{eq:(2.5)} local,
$\xi\to\xi(x)$. Then, the variation of the action~\eqref{eq:(2.1)} is given by
\begin{equation}
   \delta_\xi S=\int d^Dx\,
   \left[\partial_\mu\Bar{\xi}(x)S_\mu(x)
   -\Bar{\xi}(x)X_{\text{Fierz}}(x)\right],
\label{eq:(2.12)}
\end{equation}
where the classical supercurrent is given by
\begin{equation}
   S_\mu(x)\equiv-\frac{1}{2g_0}\sigma_{\rho\sigma}\gamma_\mu
   \psi^a(x)F_{\rho\sigma}^a(x),
\label{eq:(2.13)}
\end{equation}
and
\begin{equation}
   X_{\text{Fierz}}(x)
   \equiv\frac{1}{2}g_0
   f^{abc}\gamma_\mu\psi^a(x)\Bar{\psi}^b(x)\gamma_\mu\psi^c(x).
\label{eq:(2.14)}
\end{equation}
This $X_{\text{Fierz}}(x)$ identically vanishes for~$D=4$ because of the Fierz
identity. For~$D\neq4$, however, $X_{\text{Fierz}}(x)$ does not vanish and we
will see that in quantum theory, because of the UV divergences, this breaking
term gives rise to a nonzero contribution even for~$D\to4$.

We assume that the Faddeev--Popov ghost and anti-ghost are not transformed
under supersymmetry. Then we have the following breaking terms from the
gauge-fixing term and the ghost term:
\begin{equation}
   \delta_\xi\left(S_{\text{gf}}+S_{c\Bar{c}}\right)
   =-\int d^Dx\,\Bar{\xi}(x)\left[X_{\text{gf}}(x)+X_{c\Bar{c}}(x)\right],
\label{eq:(2.15)}
\end{equation}
where
\begin{align}
   X_{\text{gf}}(x)
   &\equiv\frac{\lambda_0}{g_0}\partial_\mu\partial_\nu A_\nu^a(x)
   \gamma_\mu\psi^a(x),
\label{eq:(2.16)}
\\
   X_{c\Bar{c}}(x)
   &\equiv\frac{1}{g_0}f^{abc}\partial_\mu\Bar{c}^a(x)c^b(x)
   \gamma_\mu\psi^c(x).
\label{eq:(2.17)}
\end{align}
We note that $X_{\text{gf}}(x)+X_{c\Bar{c}}(x)$ is BRS exact:
\begin{equation}
   X_{\text{gf}}(x)+X_{c\Bar{c}}(x)
   =\Hat{\delta}_B\frac{1}{g_0}\partial_\mu\Bar{c}^a(x)\gamma_\mu\psi^a(x)
\label{eq:(2.18)}
\end{equation}
and thus $X_{\text{gf}}(x)+X_{c\Bar{c}}(x)$ does not contribute in correlation
functions of gauge invariant operators.

Now, taking the functional integrals (here $d\mu$ denotes the functional
measure of all field variables),
\begin{equation}
   \int d\mu\,e^{-S-S_{\text{gf}}-S_{c\Bar{c}}}\,A_\alpha^b(y)\Bar{\psi}^c(z),
\label{eq:(2.19)}
\end{equation}
and
\begin{equation}
   \int d\mu\,e^{-S-S_{\text{gf}}-S_{c\Bar{c}}}
   \,\Bar{\psi}^b(y)c^c(z)\Bar{c}^d(w),
\label{eq:(2.20)}
\end{equation}
and considering the change of integration variables of the above form, we have
the following Ward--Takahashi relations:
\begin{align}
   &\left\langle
   \left[\partial_\mu S_\mu(x)+X_{\text{Fierz}}(x)
   +X_{\text{gf}}(x)+X_{c\Bar{c}}(x)\right]
   A_\alpha^b(y)\Bar{\psi}^c(z)\right\rangle
\notag\\
   &=-\delta(x-y)
   \left\langle g_0\gamma_\alpha\psi^b(y)\Bar{\psi}^c(z)\right\rangle
   -\delta(x-z)
   \left\langle A_\alpha^b(y)
   \frac{1}{2g_0}\sigma_{\beta\gamma}F_{\beta\gamma}^c(z)\right\rangle,
\label{eq:(2.21)}
\end{align}
and
\begin{align}
   &\left\langle
   \left[\partial_\mu S_\mu(x)+X_{\text{Fierz}}(x)
   +X_{\text{gf}}(x)+X_{c\Bar{c}}(x)\right]
   \Bar{\psi}^b(y)c^c(z)\Bar{c}^d(w)\right\rangle
\notag\\
   &=-\delta(x-y)
   \left\langle\frac{1}{2g_0}\sigma_{\beta\gamma}F_{\beta\gamma}^b(y)
   c^c(z)\Bar{c}^d(w)\right\rangle,
\label{eq:(2.22)}
\end{align}
where we have used the fact that the ghost and the anti-ghost are not
transformed under the super transformation. These are identities holding under
dimensional regularization.

\subsection{The effect of~$X_{\text{Fierz}}(x)$}
\begin{figure}[ht]
\begin{center}
\includegraphics[width=0.21\columnwidth]{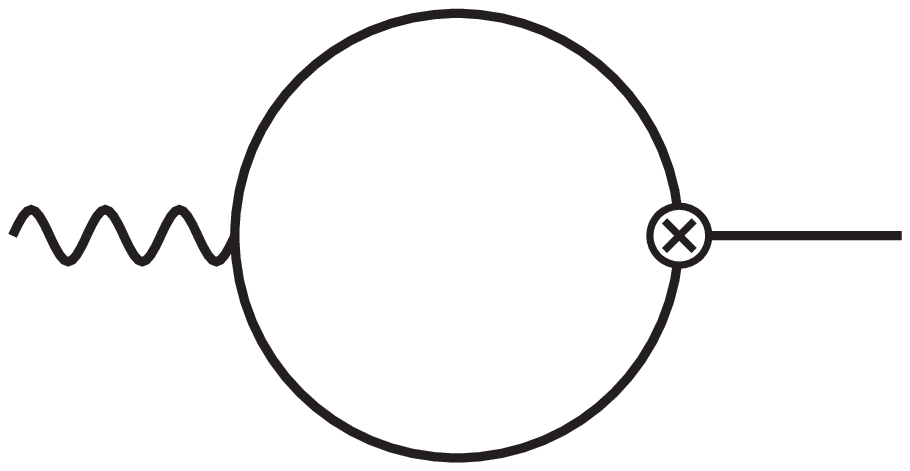}
\caption{Diagram i}
\label{fig:1}
\end{center}
\end{figure}

Let us first examine the effect of the supersymmetry breaking
term~$X_{\text{Fierz}}(x)$~\eqref{eq:(2.14)}. Since it vanishes
for~$\epsilon\to0$ (where $D=4-2\epsilon$) in the classical theory, the term
can contribute only if it is multiplied by $1/\epsilon$, i.e., only through UV
divergences in loop diagrams. The unique one-loop 1PI diagram that contributes
to the left-hand side of the WT relation~\eqref{eq:(2.21)} is diagram~i
in~Fig.~\ref{fig:1}.\footnote{In what follows, the wavy line stands for the
gauge boson propagator, the solid line the gaugino propagator, and the broken
line the ghost propagator; the blob denotes the composite operator under
consideration.} Also for the left-hand side of~Eq.~\eqref{eq:(2.22)} only
diagram~i in~Fig.~\ref{fig:1} contributes in the one-loop level because the
gaugino fields should make a loop; in~Eq.~\eqref{eq:(2.22)}, the ghost fields
are simply spectators with respect to~$X_{\text{Fierz}}(x)$.
The computation of diagram~i yields
\begin{equation}
   X_{\text{Fierz}}(x)
   \stackrel{D\to4}{\to}
   \frac{g_0}{(4\pi)^2}C_2(G)\frac{2}{3}
   \partial_\mu F_{\mu\nu}^a(x)\gamma_\nu\psi^a(x).
\label{eq:(2.23)}
\end{equation}
This tells that we have to add the following \emph{finite\/} counterterm~$S'$
to the original action~$S$,
\begin{equation}
   S'=-\frac{1}{(4\pi)^2}C_2(G)\frac{1}{6}\int d^Dx\,
   F_{\mu\nu}^a(x)F_{\mu\nu}^a(x),
\label{eq:(2.24)}
\end{equation}
so that the supervariation of this term compensates the effect
of~$X_{\text{Fierz}}(x)$ (up to $\mathcal{O}(A_\mu^2)$~terms). We thus have
modified WT relations,
\begin{align}
   &\left\langle
   \left[\partial_\mu S_\mu(x)
   +X_{\text{gf}}(x)+X_{c\Bar{c}}(x)\right]
   A_\alpha^b(y)\Bar{\psi}^c(z)\right\rangle'
\notag\\
   &=-\delta(x-y)
   \left\langle g_0\gamma_\alpha\psi^b(y)\Bar{\psi}^c(z)\right\rangle'
   -\delta(x-z)
   \left\langle A_\alpha^b(y)
   \frac{1}{2g_0}\sigma_{\beta\gamma}F_{\beta\gamma}^c(z)\right\rangle',
\label{eq:(2.25)}
\end{align}
and
\begin{align}
   &\left\langle
   \left[\partial_\mu S_\mu(x)
   +X_{\text{gf}}(x)+X_{c\Bar{c}}(x)\right]
   \Bar{\psi}^b(y)c^c(z)\Bar{c}^d(w)\right\rangle'
\notag\\
   &=-\delta(x-y)
   \left\langle\frac{1}{2g_0}\sigma_{\beta\gamma}F_{\beta\gamma}^b(y)
   c^c(z)\Bar{c}^d(w)\right\rangle'.
\label{eq:(2.26)}
\end{align}
In these expressions, the prime ($'$) implies that expectation values are
evaluated with respect to the modified action~$S+S_{\text{gf}}+S_{c\Bar{c}}+S'$,
where $S'$ is given by~Eq.~\eqref{eq:(2.24)}. The effect
of~$X_{\text{Fierz}}(x)$ in dimensional regularization is absorbed in this way.

\subsection{Renormalization of composite operators}
Next, we study the renormalization/mixing of the composite operators appearing
in~Eqs.~\eqref{eq:(2.25)} and~\eqref{eq:(2.26)}. In what follows, we adopt the
MS scheme and set
\begin{equation}
   \Delta\equiv\frac{g^2}{(4\pi)^2}C_2(G)\frac{1}{\epsilon},
\label{eq:(2.27)}
\end{equation}
where $g$ is the renormalized gauge coupling. In the Feynman gauge
$\lambda_0=1$, in the one-loop order, bare and renormalized quantities in the
$\mathcal{N}=1$ SYM in the Wess--Zumino gauge are related as (in what follows,
quantities without the subscript~$0$ and quantities with the subscript $R$
denote renormalized quantities)
\begin{align}
   g_0&=\mu^\epsilon\left(1-\frac{3}{2}\Delta\right)g,
\label{eq:(2.28)}
\\
   \lambda_0&=\left(1-\Delta\right)\lambda,
\label{eq:(2.29)}
\\
   A_\mu^a(x)
   &=\left(1-\Delta\right)A_{\mu R}^a(x),
\label{eq:(2.30)}
\\
   \begin{Bmatrix}
   \psi^a(x)\\
   \Bar{\psi}^a(x)\\
   \end{Bmatrix}
   &=\left(1-\frac{1}{2}\Delta\right)
   \begin{Bmatrix}
   \psi_R^a(x),\\
   \Bar{\psi}_R^a(x)\\
   \end{Bmatrix},
\label{eq:(2.31)}
\\
   \begin{Bmatrix}
   c^a(x)\\
   \Bar{c}^a(x)\\
   \end{Bmatrix}
   &=
   \left(1-\frac{5}{4}\Delta\right)
   \begin{Bmatrix}
   c_R^a(x)\\
   \Bar{c}_R^a(x)\\
   \end{Bmatrix},
\label{eq:(2.32)}
\end{align}
where $\mu$ is the renormalization scale and
\begin{equation}
   F_{\mu\nu}^a(x)
   =\left(1-\frac{5}{2}\Delta\right)
   \left[
   \partial_\mu A_{\nu R}^a(x)-\partial_\nu A_{\mu R}^a(x)\right]
   +\left(1-\frac{11}{4}\Delta\right)
   \{f^{abc}A_\mu^bA_\nu^c\}_R(x).
\label{eq:(2.33)}
\end{equation}
The last term denotes the renormalized composite operator corresponding
to~$f^{abc}A_\mu^b(x)A_\nu^c(x)$.

\begin{figure}[ht]
\begin{minipage}{0.3\columnwidth}
\begin{center}
\includegraphics[width=0.7\columnwidth]{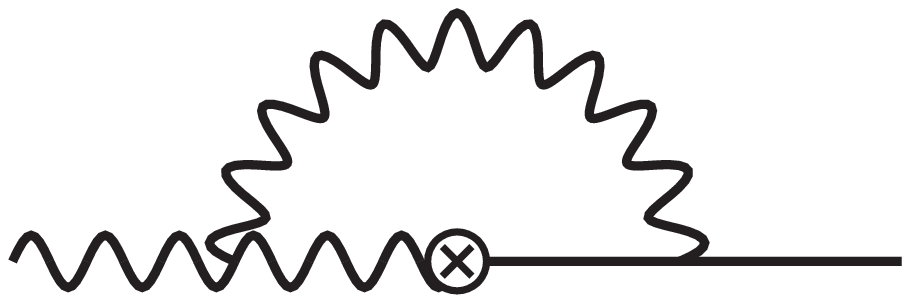}
\caption{Diagram a}
\label{fig:2}
\end{center}
\end{minipage}
\begin{minipage}{0.3\columnwidth}
\begin{center}
\includegraphics[width=0.7\columnwidth]{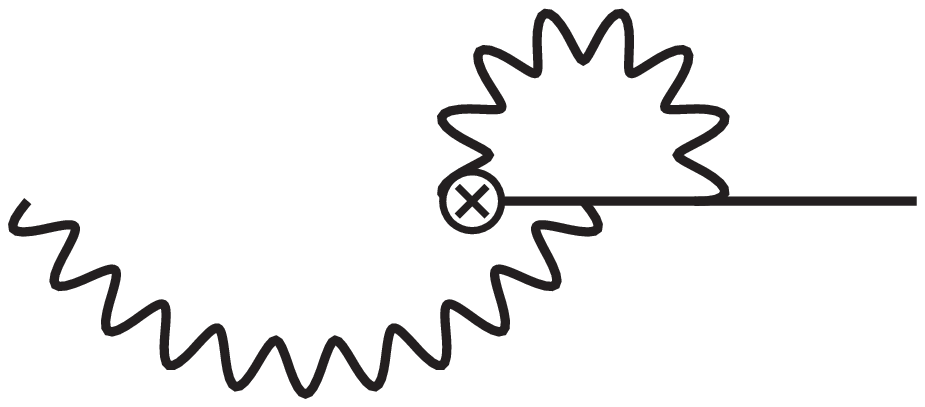}
\caption{Diagram b}
\label{fig:3}
\end{center}
\end{minipage}
\begin{minipage}{0.3\columnwidth}
\begin{center}
\includegraphics[width=0.7\columnwidth]{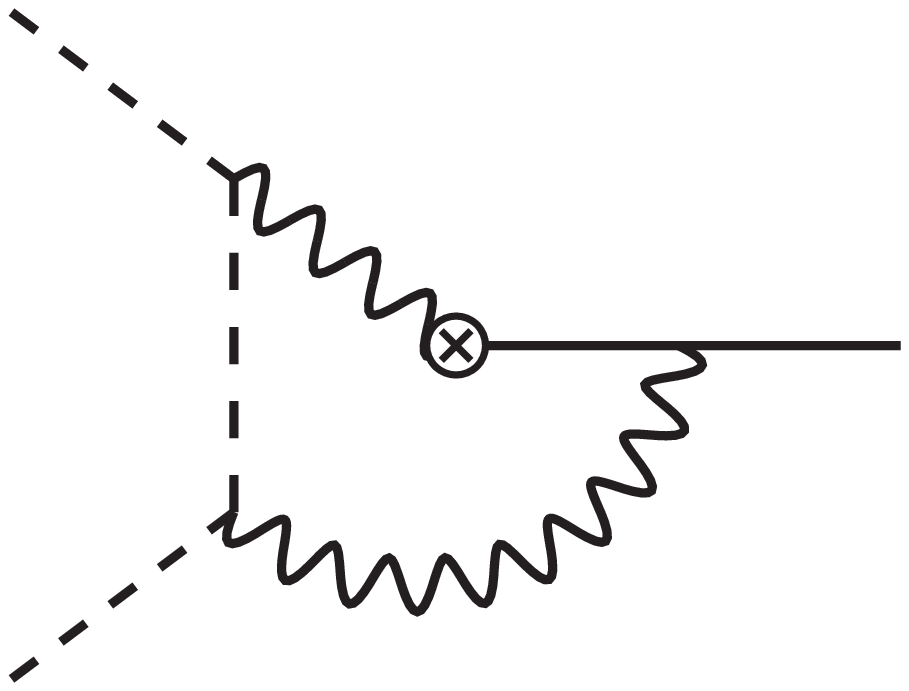}
\caption{Diagram c}
\label{fig:4}
\end{center}
\end{minipage}
\end{figure}

\begin{figure}[ht]
\begin{minipage}{0.3\columnwidth}
\begin{center}
\includegraphics[width=0.7\columnwidth]{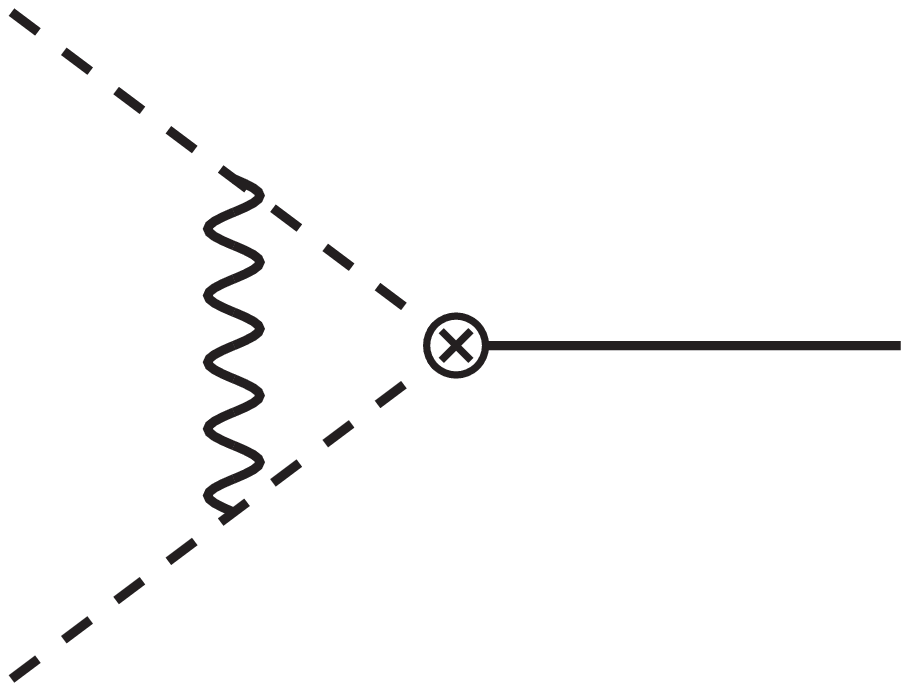}
\caption{Diagram d}
\label{fig:5}
\end{center}
\end{minipage}
\begin{minipage}{0.3\columnwidth}
\begin{center}
\includegraphics[width=0.7\columnwidth]{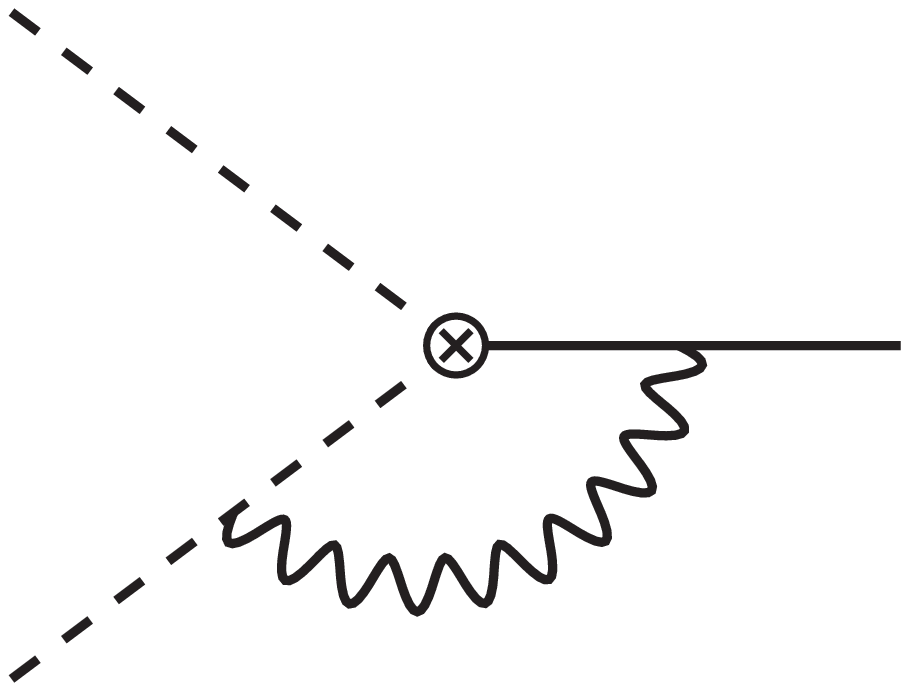}
\caption{Diagram e}
\label{fig:6}
\end{center}
\end{minipage}
\begin{minipage}{0.3\columnwidth}
\begin{center}
\includegraphics[width=0.7\columnwidth]{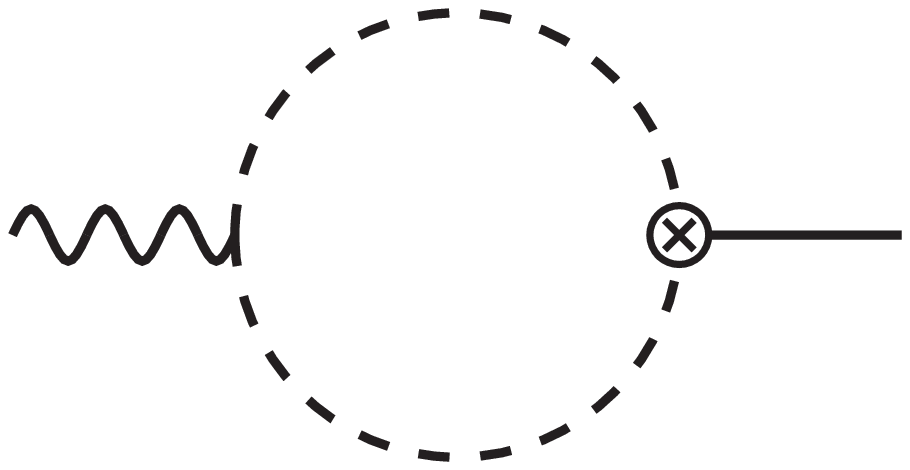}
\caption{Diagram f}
\label{fig:7}
\end{center}
\end{minipage}
\end{figure}

Let us next consider the renormalization of the composite operators
$X_{\text{gf}}(x)$ in~Eq.~\eqref{eq:(2.16)} and~$X_{c\Bar{c}}(x)$
in~Eq.~\eqref{eq:(2.17)}. By substituting bare quantities
in~Eqs.~\eqref{eq:(2.16)} and~\eqref{eq:(2.17)}
by~Eqs.~\eqref{eq:(2.28)}--\eqref{eq:(2.32)}, in the one-loop order, we have
\begin{align}
   X_{\text{gf}}(x)
   &=(1-\Delta)\frac{\lambda}{g}\partial_\mu\partial_\nu A_{\nu R}^a(x)
   \gamma_\mu\psi_R^a(x),
\label{eq:(2.34)}
\\
   X_{c\Bar{c}}(x)
   &=\left(1-\frac{3}{2}\Delta\right)
   \frac{1}{g}f^{abc}\partial_\mu\Bar{c}_R^a(x)c_R^b(x)
   \gamma_\mu\psi_R^c(x).
\label{eq:(2.35)}
\end{align}
The UV divergences (being proportional to~$\Delta$) in these expressions arise
from self-energy corrections in external lines of the composite operators and
the renormalization of involved parameters. Besides these divergences, the
composite operators produce further divergences associated with the vertex part
(i.e., 1PI part) containing the composite operators. Relevant one-loop diagrams
are depicted in~Figs.~\ref{fig:2}--\ref{fig:7}.\footnote{The
counterterm~$S'$ in~Eq.~\eqref{eq:(2.24)} does not influence the following
one-loop analysis of the renormalization of composite operators, because $S'$
is a one-loop order quantity and it is UV finite.}

In the Feynman gauge $\lambda_0=1$, we find that the sum of divergent parts of
these diagrams is given by
\begin{align}
   &\left.\left[
   \frac{\lambda}{g}\partial_\mu\partial_\nu A_{\nu R}^a(x)
   \gamma_\mu\psi_R^a(x)
   +\frac{1}{g}f^{abc}\partial_\mu\Bar{c}_R^a(x)c_R^b(x)\gamma_\mu\psi_R^c(x)
   \right]
   \right|_{\text{1PI, divergent part}}
\notag\\
   &=2\Delta
   \frac{\lambda}{g}\partial_\mu\partial_\nu A_{\nu R}^a(x)
   \gamma_\mu\psi_R^a(x)
   +\frac{1}{2}\Delta
   \frac{1}{g}f^{abc}\partial_\mu\Bar{c}_R^a(x)c_R^b(x)\gamma_\mu\psi_R^c(x)
\notag\\
   &\qquad{}
   +\Delta
   \partial_\mu\left\{
   -\frac{1}{2g}\sigma_{\rho\sigma}\gamma_\mu
   \psi_R^a(x)\left[
   \partial_\rho A_{\sigma R}^a(x)-\partial_\sigma A_{\rho R}^a(x)
   \right]\right\}
\notag\\
   &\qquad\qquad{}
   +2\Delta\left(-\frac{1}{g^2}\right)
   \partial_\mu\partial_\mu A_{\nu R}^a(x)g\gamma_\nu\psi_R^a(x)
\notag\\
   &\qquad\qquad\qquad{}
   +\frac{3}{2}\Delta\frac{1}{2g}
   \left[
   \partial_\mu A_{\nu R}^a(x)-\partial_\nu A_{\mu R}^a(x)
   \right]\sigma_{\mu\nu}
   \Slash{\partial}\psi_R^a(x)
\notag\\
   &\qquad\qquad\qquad\qquad{}
   +\Delta\frac{1}{4g}\partial_\mu
   \left\{\left[A_{\nu R}^a(x)\gamma_\nu\gamma_\mu+2A_{\mu R}^a(x)\right]
   \Slash{\partial}\psi_R^a(x)\right\}
   +\Delta\mathcal{O}(A_{\mu R}^2).
\label{eq:(2.36)}
\end{align}
Since here we are considering only correlation functions with at most
one~$A_\mu$-line, we cannot determine the last
$\Delta\mathcal{O}(A_{\mu R}^2)$~term; we will fix the associated ambiguity in
the supercurrent by imposing gauge invariance in the very final stage of
our analysis.

Then, from the sum of~Eqs.~\eqref{eq:(2.34)}, \eqref{eq:(2.35)},
and~\eqref{eq:(2.36)}, we see that to one-loop order all the divergences
are canceled out in~$X_{\text{gf}R}(x)$ and in~$X_{c\Bar{c}R}(x)$ defined by
\begin{align}
   &X_{\text{gf}}(x)+X_{c\Bar{c}}(x)
\notag\\
   &\equiv(1+\Delta)X_{\text{gf}R}(x)+(1-\Delta)X_{c\Bar{c}R}(x)
\notag\\
   &\qquad{}
   +\Delta
   \partial_\mu\left\{
   -\frac{1}{2g}\sigma_{\rho\sigma}\gamma_\mu
   \psi_R^a(x)\left[\partial_\rho A_{\sigma R}^a(x)
   -\partial_\sigma A_{\rho R}^a(x)
   \right]\right\}
\notag\\
   &\qquad\qquad{}
   +2\Delta\left(-\frac{1}{g^2}\right)
   \partial_\mu\partial_\mu A_{\nu R}^a(x)g\gamma_\nu\psi_R^a(x)
\notag\\
   &\qquad\qquad\qquad{}
   +\frac{3}{2}\Delta\frac{1}{2g}
   \left[
   \partial_\mu A_{\nu R}^a(x)-\partial_\nu A_{\mu R}^a(x)
   \right]\sigma_{\mu\nu}
   \Slash{\partial}\psi_R^a(x)
\notag\\
   &\qquad\qquad\qquad\qquad{}
   +\Delta\frac{1}{4g}\partial_\mu
   \left\{\left[A_{\nu R}^a(x)\gamma_\nu\gamma_\mu
   +2A_{\mu R}^a(x)\right]\Slash{\partial}\psi_R^a(x)\right\}
   +\Delta\mathcal{O}(A_{\mu R}^2).
\label{eq:(2.37)}
\end{align}
Here, we have assumed that $X_{\text{gf}R}(x)$ and~$X_{c\Bar{c}R}(x)$ in the tree
level are given by bare ones, $X_{\text{gf}}(x)$ and~$X_{c\Bar{c}}(x)$,
respectively. Thus, $X_{\text{gf}R}(x)$ and~$X_{c\Bar{c}R}(x)$ in the above
expression are renormalized finite composite operators corresponding to
$X_{\text{gf}}(x)$ and~$X_{c\Bar{c}}(x)$, respectively.

\begin{figure}[ht]
\begin{minipage}{0.3\columnwidth}
\begin{center}
\includegraphics[width=0.7\columnwidth]{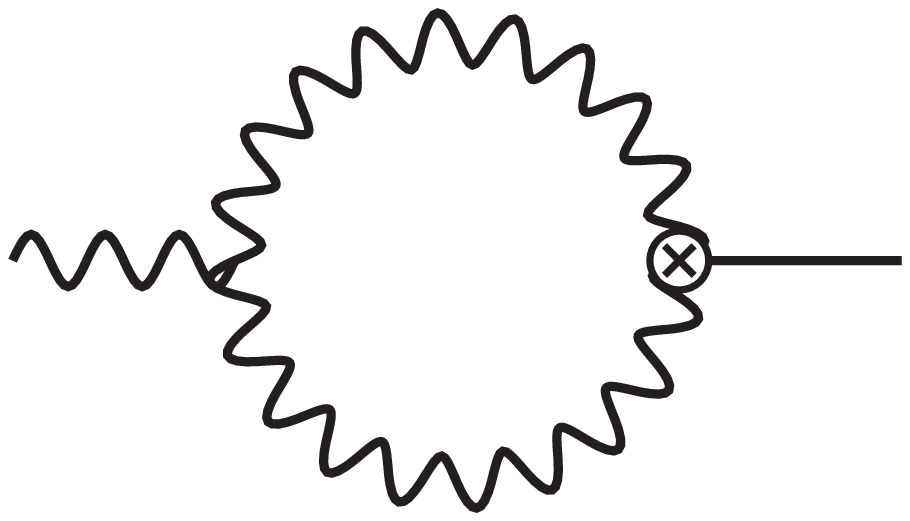}
\caption{Diagram g}
\label{fig:8}
\end{center}
\end{minipage}
\begin{minipage}{0.3\columnwidth}
\begin{center}
\includegraphics[width=0.7\columnwidth]{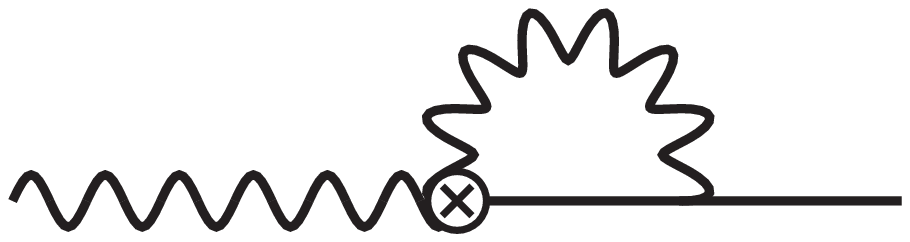}
\caption{Diagram h}
\label{fig:9}
\end{center}
\end{minipage}
\end{figure}

Next, we analyze the supercurrent~$S_\mu(x)$ in~Eq.~\eqref{eq:(2.13)}. The
substitutions~\eqref{eq:(2.28)}, \eqref{eq:(2.30)}, and~\eqref{eq:(2.31)} give
rise to
\begin{equation}
   S_\mu(x)=-\frac{1}{2g}\sigma_{\rho\sigma}\gamma_\mu\psi_R^a(x)
   \left[\partial_\rho A_{\sigma R}^a(x)-\partial_\sigma A_{\rho R}^a(x)
   +f^{abc}A_{\rho R}^b(x)A_{\sigma R}^c(x)\right]+\Delta\mathcal{O}(A_{\mu R}^2).
\label{eq:(2.38)}
\end{equation}
Here, note that the coefficient of the $f^{abc}A_{\rho R}^b(x)A_{\sigma R}^c(x)$
term is uniquely fixed  because the $\Delta\mathcal{O}(A_{\mu R}^2$ term should
be proportional to~$\Delta$. On the other hand, a somewhat tedious calculation
of diagrams in~Figs.~\ref{fig:2}, \ref{fig:3}, \ref{fig:8}, and~\ref{fig:9}
shows that\footnote{One-loop diagrams with external ghost lines turn out to be
UV finite.}
\begin{align}
   &\left.-\frac{1}{2g}\sigma_{\rho\sigma}\gamma_\mu\psi_R^a(x)
   \left[\partial_\rho A_{\sigma R}^a(x)-\partial_\sigma A_{\rho R}^a(x)
   +f^{abc}A_{\rho R}^b(x)A_{\sigma R}^c(x)\right]\right|_{\text{1PI, divergent part}}
\notag\\
   &=-\Delta\frac{1}{4g}
   \left[A_{\nu R}^a(x)\gamma_\nu\gamma_\mu
   +2A_{\mu R}^a(x)\right]\Slash{\partial}\psi_R^a(x)
   +\Delta\mathcal{O}(A_{\mu R}^2).
\label{eq:(2.39)}
\end{align}
Then from the sum of~Eqs.~\eqref{eq:(2.38)} and~\eqref{eq:(2.39)}, we see that
all divergences are canceled out in the combination~$S_{\mu R}(x)$, defined by
\begin{equation}
   S_\mu(x)
   \equiv S_{\mu R}(x)
   -\Delta\frac{1}{4g}
   \left[A_{\nu R}^a(x)\gamma_\nu\gamma_\mu
   +2A_{\mu R}^a(x)\right]\Slash{\partial}\psi_R^a(x)
   +\Delta\mathcal{O}(A_{\mu R}^2).
\label{eq:(2.40)}
\end{equation}
Thus, $S_{\mu R}(x)$ is the renormalized supercurrent up to a one-loop
$\mathcal{O}(A_{\mu R}^2)$ term.

In summary, the renormalization of the composite operators, $X_{\text{gf}}(x)$,
$X_{c\Bar{c}}(x)$, and~$S_\mu(x)$, is accomplished by~Eqs.~\eqref{eq:(2.37)}
and~\eqref{eq:(2.40)}; $X_{\text{gf}R}(x)$,
$X_{c\Bar{c}R}(x)$, and~$S_{\mu R}(x)$ are corresponding renormalized composite
operators.

\subsection{Supersymmetry WT relations in terms of renormalized operators}
Now, we substitute Eqs.~\eqref{eq:(2.28)}--\eqref{eq:(2.33)},
\eqref{eq:(2.37)}, and~\eqref{eq:(2.40)} in the WT identities
Eqs.~\eqref{eq:(2.25)} and~\eqref{eq:(2.26)}. For Eq.~\eqref{eq:(2.25)}, we
note the following relations hold in the \emph{tree-level approximation\/}
(in the Feynman gauge $\lambda_0=1$):
\begin{align}
   \left\langle
   \left(-\frac{1}{g^2}\right)
   \partial_\mu\partial_\mu A_{\nu R}^a(x)A_{\rho R}^b(y)\right\rangle'
   &=\delta^{ab}\delta_{\nu\rho}\delta(x-y),
\label{eq:(2.41)}
\\
   \left\langle\Slash{\partial}\psi_R^a(x)\Bar{\psi}_R^b(y)\right\rangle'
   &=\delta^{ab}\delta(x-y),
\label{eq:(2.42)}
\\
   \left\langle X_{c\Bar{c} R}(x)A_{\alpha R}^b(y)\Bar{\psi}_R^c(z)\right\rangle'
   &=0.
\label{eq:(2.43)}
\end{align}
We can use these tree-level relations in the terms proportional to~$\Delta$,
because $\Delta$ is already a one-loop-order quantity.\footnote{We
also note that we can neglect the effect of the
counterterm~$S'$ in~Eq.~\eqref{eq:(2.24)}
to~Eqs.~\eqref{eq:(2.41)}--\eqref{eq:(2.43)}, because Eq.~\eqref{eq:(2.24)} is
already a one-loop-order quantity and we use
Eqs.~\eqref{eq:(2.41)}--\eqref{eq:(2.43)} in the terms proportional
to~$\Delta$.} Then we find, to the one-loop order,
\begin{align}
   &\left\langle
   \left[
   \partial_\mu S_{\mu R}(x)+X_{\text{gf}R}(x)+X_{c\Bar{c}R}(x)
   +\Delta\mathcal{O}(A_{\mu R}^2)\right]
   A_{\alpha R}^b(y)\Bar{\psi}_R^c(z)\right\rangle'
\notag\\
   &=-\delta(x-y)
   \left\langle g\gamma_\alpha\psi_R^b(y)\Bar{\psi}_R^c(z)\right\rangle'
\notag\\
   &\qquad{}
   -\delta(x-z)
   \left\langle A_{\alpha R}^b(y)
   \frac{1}{2g}\sigma_{\beta\gamma}
   \left[
   \partial_\beta A_{\gamma R}^c(z)-\partial_\gamma A_{\beta R}^c(z)
   +\{f^{cde}A_\beta^dA_\gamma^e\}_R(z)
   \right]
   \right\rangle'.
\label{eq:(2.44)}
\end{align}

For Eq.~\eqref{eq:(2.26)}, on the other hand, by using the tree-level
relation
\begin{align}
   &\left\langle
   \left(-\frac{1}{g^2}\right)
   \partial_\mu\partial_\mu A_{\nu R}^a(x)g\gamma_\nu\psi_R^a(x)
   \Bar{\psi}_R^b(y)c_R^c(z)\Bar{c}_R^d(w)\right\rangle'
\notag\\
   &=\left\langle X_{c\Bar{c}}(x)
   \Bar{\psi}_R^b(y)c_R^c(z)\Bar{c}_R^d(w)\right\rangle',
\label{eq:(2.45)}
\end{align}
in the term proportional to~$\Delta$, we have in the one-loop order,
\begin{align}
   &\left\langle
   \left[\partial_\mu S_{\mu R}(x)+X_{\text{gf}R}(x)+X_{c\Bar{c} R}(x)
   +\Delta\mathcal{O}(A_{\mu R}^2)\right]
   \Bar{\psi}_R^b(y)c_R^c(z)\Bar{c}_R^d(w)\right\rangle'
\notag\\
   &=-\delta(x-y)
   \left\langle\frac{1}{2g}\sigma_{\beta\gamma}
   \left[
   \partial_\beta A_{\gamma R}^b(y)-\partial_\gamma A_{\beta R}^b(y)
   +\{f^{bef}A_\beta^eA_\gamma^f\}_R(y)
   \right]
   c_R^c(z)\Bar{c}_R^d(w)\right\rangle'.
\label{eq:(2.46)}
\end{align}

Remarkably, Eqs.~\eqref{eq:(2.44)} and~\eqref{eq:(2.46)} show that in the
one-loop order, the following combination of renormalized finite composite
operators,
\begin{equation}
   \partial_\mu S_{\mu R}(x)+X_{\text{gf}R}(x)+X_{c\Bar{c}R}(x),
\label{eq:(2.47)}
\end{equation}
up to possible $\Delta\mathcal{O}(A_{\mu R}^2)$ terms that cannot be read off
from our present calculation, generates \emph{the properly normalized super
transformation on renormalized elementary fields}. The existence of such a
finite operator would be expected on general grounds (i.e., supersymmetry
should be free from quantum anomaly). Nevertheless, the validity of
supersymmetry WT relations in the Wess--Zumino gauge that we have observed
above still appears miraculous, because it resulted from nontrivial
renormalization/mixing of various composite operators.

\subsection{Final step}
We have observed that the combination~\eqref{eq:(2.47)} generates the correct
super transformation on renormalized elementary fields. Whether the same
combination generates the correct super transformation on \emph{composite
operators\/} is far from obvious and to answer this requires further
complicated analyses. However, if we consider only ``on-shell'' correlation
functions in which all composite operators are separated from the
combination~\eqref{eq:(2.47)} in position space, we can still regard the
combination~\eqref{eq:(2.47)} as properly normalized because new UV divergences
associated with composite operators at an equal point do not arise.

Within such on-shell correlation functions, in the one-loop order we can
neglect the term proportional to~$\Delta\Slash{\partial}\psi_R(x)$ in the
supercurrent~\eqref{eq:(2.40)}, because one can use the tree-level equation of
motion of the gaugino field. Moreover, since we are practically interested only
in correlation functions of gauge-invariant operators, requiring that the
supercurrent is gauge invariant, we can eliminate the possibility of the
$\Delta\mathcal{O}(A_{\mu R}^2)$ term in~Eq.~\eqref{eq:(2.40)}
(note that there is no dimension~$7/2$ gauge-invariant fermionic combination
of~$\mathcal{O}(A_\mu^2)$). Thus, in the one-loop order, we may set
\begin{equation}
   S_{\mu R}(x)\to
   S_\mu(x)
   =-\frac{1}{2g_0}\sigma_{\rho\sigma}\gamma_\mu
   \psi^a(x)F_{\rho\sigma}^a(x),
\label{eq:(2.48)}
\end{equation}
in the on-shell correlation functions of gauge-invariant operators.

For the combination $X_{\text{gf}R}(x)+X_{c\Bar{c}R}(x)$ in~Eq.~\eqref{eq:(2.47)},
by using the identity at~$D=4$,
\begin{equation}
   \sigma_{\rho\sigma}\gamma_\mu
   =\gamma_\rho\delta_{\sigma\mu}-\gamma_\sigma\delta_{\rho\mu}
   -\gamma_5\gamma_\alpha\epsilon_{\alpha\rho\sigma\mu},
\label{eq:(2.49)}
\end{equation} 
we can rewrite Eq.~\eqref{eq:(2.37)} as
\begin{align}
   X_{\text{gf}R}(x)+X_{c\Bar{c}R}(x)
   &=X_{\text{gf}}(x)+X_{c\Bar{c}}(x)
   +\Delta(\text{terms proportional to
   $\partial_\mu\partial_\mu A_\nu(x)$ or $\Slash{\partial}\psi(x)$})
\notag\\
   &\qquad{}
   +\Delta\frac{1}{g_0}f^{abc}\partial_\mu\Bar{c}^a(x)c^b(x)\gamma_\mu\psi^c(x)
   +\Delta\mathcal{O}(A_{\mu R}^2)
\label{eq:(2.50)}
\end{align}
in the one-loop order; here we have used the Feynman gauge~$\lambda_0=1$. For
the insertion of this combination in the on-shell correlation functions of
gauge invariant operators, the first line of the right-hand side can be
neglected because of the BRS exactness~\eqref{eq:(2.18)}\footnote{The presence
of the counterterm~$S'$ in~Eq.~\eqref{eq:(2.24)} does not influence this
argument, because $S'$ is invariant under the BRS
transformation~\eqref{eq:(2.9)}.} and the tree-level equations of motion in the
Feynman gauge. If we consider only operators that do not contain the ghost
field and the anti-ghost field, the term
$\Delta f^{abc}\partial_\mu\Bar{c}^a(x)c^b(x)\gamma_\mu\psi^c(x)$ can also
be neglected because there is no tree-level diagram in which this term can
contribute.

From these considerations, for the insertion of the
combination~\eqref{eq:(2.47)} in the on-shell correlation functions of
gauge-invariant operators that do not contain the ghost field and the
anti-ghost field, we can set
\begin{equation}
   \partial_\mu S_{\mu R}(x)+X_{\text{gf}R}(x)+X_{c\Bar{c}R}(x)
   \to\partial_\mu S_\mu(x)+\Delta\mathcal{O}(A_{\mu R}^2)
   \to\partial_\mu S_\mu(x),
\label{eq:(2.51)}
\end{equation}
in the one-loop order; in the last step, we have invoked gauge invariance of
the whole expression.

The bottom line of the above lengthy analysis is that, under dimensional
regularization, the properly normalized conserved supercurrent in (on-shell)
correlation functions of gauge-invariant operators (which do not contain the
ghost field and the anti-ghost field) is given by
\begin{equation}
   S_{\mu R}(x)
   =-\frac{1}{2g_0}\sigma_{\rho\sigma}\gamma_\mu
   \psi^a(x)F_{\rho\sigma}^a(x)+\mathcal{O}(g_0^3).
\label{eq:(2.52)}
\end{equation}
This is the result that we have already announced in~Eq.~\eqref{eq:(1.2)}.

Having obtained the expression for the correctly normalized conserved
supercurrent~\eqref{eq:(2.52)}, in the next section we will construct a
composite operator of the flowed fields that reproduces Eq.~\eqref{eq:(2.52)}
in the small flow-time limit.

\section{Representation of the supercurrent in terms of flowed fields}
\label{sec:3}
\subsection{Flow equations}
The flow equations we adopt in this paper are identical to those
of~Refs.~\cite{Luscher:2009eq,Luscher:2010iy,Luscher:2013cpa}. The flow of the
gauge field along the flow time~$t\geq0$ is defined by
\begin{equation}
   \partial_tB_\mu(t,x)=D_\nu G_{\nu\mu}(t,x),\qquad
   B_\mu(t=0,x)=A_\mu(x),
\label{eq:(3.1)}
\end{equation}
where
\begin{equation}
   D_\mu\equiv\partial_\mu+[B_\mu,\cdot],\qquad
   G_{\mu\nu}(t,x)
   \equiv\partial_\mu B_\nu(t,x)-\partial_\nu B_\mu(t,x)
   +[B_\mu(t,x),B_\nu(t,x)],
\label{eq:(3.2)}
\end{equation}
and the flow for the fermion (gaugino) fields is defined by
\begin{align}
   \partial_t\chi(t,x)&=\mathcal{D}^2\chi(t,x),&
   \chi(t=0,x)&=\psi(x),
\label{eq:(3.3)}
\\
   \partial_t\Bar{\chi}(t,x)
   &=\Bar{\chi}(t,x)
   \overleftarrow{\mathcal{D}}^2,&
   \Bar{\chi}(t=0,x)&=\Bar{\psi}(x),
\label{eq:(3.4)}
\end{align}
where the covariant derivatives in the adjoint representation are defined by
\begin{align}
   \mathcal{D}_\mu^{ab}&\equiv\delta^{ab}\partial_\mu+f^{acb}B_\mu^c(t,x)
   \equiv\delta^{ab}\partial_\mu+\mathcal{B}_\mu^{ab}(t,x),
\label{eq:(3.5)}
\\
   \overleftarrow{\mathcal{D}}_\mu^{ab}
   &\equiv\delta^{ab}\overleftarrow{\partial}_\mu-f^{acb}B_\mu^c(t,x)
   \equiv\delta^{ab}\overleftarrow{\partial}_\mu-\mathcal{B}_\mu^{ab}(t,x).
\label{eq:(3.6)}
\end{align}
The fields, $B(t,x)$, $\chi(t,x)$, and~$\Bar{\chi}(t,x)$, are referred to as
\emph{flowed fields\/} throughout this paper.

\subsection{The small flow-time expansion of composite operators}
We want to express the composite operator~\eqref{eq:(2.52)} in the original
gauge theory in terms of the small flow-time $t\to0$ limit of flowed fields,
because of the nice renormalization property of flowed fields. This is achieved
if we can find the coefficients in the small flow-time
expansion (Ref.~\cite{Luscher:2011bx}). To be explicit, we need to know the
coefficients~$\zeta_i(t)$ ($i=1$, $2$, $3$) in the following asymptotic
expansion for~$t\to0$:
\begin{align}
   \chi^a(t,x)G_{\mu\nu}^a(t,x)
   &=\zeta_1(t)\psi^a(x)F_{\mu\nu}^a(x)
\notag\\
   &\qquad{}
   +\zeta_2(t)\left[
   \gamma_\mu\gamma_\rho\psi^a(x)F_{\rho\nu}^a(x)
   -\gamma_\nu\gamma_\rho\psi^a(x)F_{\rho\mu}^a(x)
   \right]
\notag\\
   &\qquad\qquad{}
   +\zeta_3(t)\sigma_{\rho\sigma}\sigma_{\mu\nu}\psi^a(x)
   F_{\rho\sigma}^a(x)+\mathcal{O}(t),
\label{eq:(3.7)}
\end{align}
where the composite operators in the right-hand side possess
the same mass dimension ($=7/2$) and the same gauge, Lorentz, and parity
structures as the left-hand side. For $t\to0$, perturbation theory is justified
owing to the asymptotic freedom (see below) and the coefficients have the
following loop expansion,
\begin{equation}
   \zeta_1(t)=1+\zeta_1^{(1)}(t)+\dotsb,\qquad
   \zeta_2(t)=\zeta_2^{(1)}(t)+\dotsb,\qquad
   \zeta_3(t)=\zeta_3^{(1)}(t)+\dotsb,
\label{eq:(3.8)}
\end{equation}
where $\zeta_i^{(1)}(t)$ are one-loop-order coefficients. Thus, to the one-loop
order, we can invert the relation~\eqref{eq:(3.7)} with respect
to~$\psi^a(x)F_{\mu\nu}^a(x)$ as
\begin{align}
   \psi^a(x)F_{\mu\nu}^a(x)
   &=\left[1-\zeta_1^{(1)}(t)\right]\chi^a(t,x)G_{\mu\nu}^a(t,x)
\notag\\
   &\qquad{}
   -\zeta_2^{(1)}(t)
   \left[
   \gamma_\mu\gamma_\rho\chi^a(t,x)G_{\rho\nu}^a(t,x)
   -\gamma_\nu\gamma_\rho\chi^a(t,x)G_{\rho\mu}^a(t,x)\right]
\notag\\
   &\qquad\qquad{}
   -\zeta_3^{(1)}(t)
   \sigma_{\rho\sigma}\sigma_{\mu\nu}\chi^a(t,x)G_{\rho\sigma}^a(t,x)
   +\mathcal{O}(t),
\label{eq:(3.9)}
\end{align}
and then to the one-loop order the supercurrent~\eqref{eq:(2.52)} is expressed
as
\begin{align}
   S_{\mu R}(x)
   &=-\frac{1}{2g_0}\sigma_{\rho\sigma}\gamma_\mu\psi^a(x)F_{\rho\sigma}^a(x)
   +\mathcal{O}(g_0^3)
\notag\\
   &=
   -\frac{1}{2g_0}\left[1-\zeta_1^{(1)}(t)-2(D-3)\zeta_2^{(1)}(t)
   +(D-9)(D-4)\zeta_3^{(1)}(t)\right]
\notag\\
   &\qquad\qquad\qquad\qquad\qquad\qquad\qquad\qquad
   \qquad\qquad{}
   \times\sigma_{\rho\sigma}\gamma_\mu\chi^a(t,x)G_{\rho\sigma}^a(t,x)
\notag\\
   &\qquad{}
   -\frac{1}{2g_0}\left[4(D-4)\zeta_2^{(1)}(t)-4(D-5)(D-4)\zeta_3^{(1)}(t)\right]
\notag\\
   &\qquad\qquad\qquad\qquad\qquad\qquad\qquad\qquad
   \qquad{}
   \times\gamma_\nu\chi^a(t,x)G_{\nu\mu}^a(t,x)+\mathcal{O}(t)
   +\mathcal{O}(g_0^3).
\label{eq:(3.10)}
\end{align}
This is the relation that we wanted to have. Thus, our next task is to
compute the one-loop expansion coefficients~$\zeta_i^{(1)}$.

\subsection{One-loop computation of expansion coefficients}
The background field method developed in~Ref.~\cite{Suzuki:2015bqa} is a
powerful method to compute the coefficients in the small flow-time expansion.
See also Ref.~\cite{Suzuki:2015kba}. In this method, we decompose fields into
the background $c$-number part and the quantum fluctuating part as
\begin{align}
   A_\mu(x)&=\Hat{A}_\mu(x)+a_\mu(x),&
   B_\mu(t,x)&=\Hat{B}_\mu(t,x)+b_\mu(t,x),
\label{eq:(3.11)}
\\
   \psi(x)&=\Hat{\psi}(x)+p(x),&
   \chi(t,x)&=\Hat{\chi}(t,x)+k(t,x),
\label{eq:(3.12)}
\\
   \Bar{\psi}(x)&=\Hat{\Bar{\psi}}(x)+\Bar{p}(x),&
   \Bar{\chi}(t,x)&=\Hat{\Bar{\chi}}(t,x)+\Bar{k}(t,x).
\label{eq:(3.13)}
\end{align}
Quantities with a hat ($\Hat{\phantom{x}}$) are background ones. Then we
introduce the background gauge-covariant gauge-fixing term
\begin{equation}
   S_{\text{gf}}
   =\frac{\lambda_0}{2g_0^2}\int d^4x\,
   \Hat{D}_\mu a_\mu^a(x)\Hat{D}_\nu a_\nu^a(x),
\label{eq:(3.14)}
\end{equation}
where $\hat{D}_\mu$ is the covariant derivative with respect
to~$\Hat{A}_\mu(x)$ instead of~Eq.~\eqref{eq:(2.7)}. We also adopt the
following ``gauge-fixed'' flow equations:
\begin{align}
   \partial_tB_\mu(t,x)&=D_\nu G_{\nu\mu}(t,x)
   +\alpha_0D_\mu\Hat{D}_\nu b_\nu(t,x),&
   B_\mu(t=0,x)&=A_\mu(x),
\label{eq:(3.15)}
\\
   \partial_t\chi^a(t,x)&
   =\left\{(\mathcal{D}^2)^{ab}-\alpha_0f^{acb}[\Hat{D}_\mu b_\mu(t,x)]^c\right\}
   \chi^b(t,x),&
   \chi^a(t=0,x)&=\psi^a(x),
\label{eq:(3.16)}
\\
   \partial_t\Bar{\chi}^a(t,x)
   &=\Bar{\chi}^b(t,x)
   \left\{(\overleftarrow{\mathcal{D}}^2)^{ba}
   +\alpha_0f^{bca}[\Hat{D}_\mu b_\mu(t,x)]^c\right\},&
   \Bar{\chi}^a(t=0,x)&=\Bar{\psi}^a(x).
\label{eq:(3.17)}
\end{align}
In what follows, we work with the ``Feynman gauge'', $\lambda_0=\alpha_0=1$.
We postulate that the background fields obey their own flow
equations (Ref.~\cite{Suzuki:2015bqa}):
\begin{align}
   \partial_t\Hat{B}_\mu(t,x)&=\Hat{D}_\nu\Hat{G}_{\nu\mu}(t,x),&
   \Hat{B}_\mu(t=0,x)&=\Hat{A}_\mu(x),
\label{eq:(3.18)}
\\
   \partial_t\Hat{\chi}(t,x)
   &=\Hat{\mathcal{D}}^2\Hat{\chi}(t,x),&
   \Hat{\chi}(t=0,x)&=\Hat{\psi}(x),
\label{eq:(3.19)}
\\
   \partial_t\Hat{\Bar{\chi}}(t,x)
   &=\Hat{\Bar{\chi}}(t,x)
   \Hat{\overleftarrow{\mathcal{D}}}^2,&
   \Hat{\Bar{\chi}}(t=0,x)&=\Hat{\Bar{\psi}}(x),
\label{eq:(3.20)}
\end{align}
where the quantities with a hat are given by corresponding quantities with the
replacement $B_\mu(t,x)\to\Hat{B}_\mu(t,x)$. We assume
$\Hat{D}_\nu\Hat{F}_{\nu\mu}(x)=0$ (Ref.~\cite{Suzuki:2015bqa}) so that
the background gauge field does not evolve, $\Hat{B}_\mu(t,x)=\Hat{A}_\mu(x)$.
We further assume that $\Hat{\Slash{\mathcal{D}}}^{ab}\Hat{\psi}^b(x)=%
\Hat{\Bar{\psi}}^a(x)\Hat{\overleftarrow{\Slash{\mathcal{D}}}}^{ab}=0$ to
suppress the tree-level tadpoles, $\langle p^a(x)\rangle^{(0)}=%
\langle\Bar{p}^a(x)\rangle^{(0)}=0$, where the superscript~$^{(0)}$ implies that
the expectation values are computed in the tree-level approximation. These
assumptions also imply $\langle a^a(x)\rangle^{(0)}=\mathcal{O}(\Hat{\psi}^2)$
and, through the flow equations of quantum fields (Ref.~\cite{Suzuki:2015bqa}),
$\langle b^a(t,x)-a^a(x)\rangle^{(0)}=\mathcal{O}(t,\Hat{\psi}^2)$
and~$\langle k^a(t,x)\rangle^{(0)}=\langle\Bar{k}^a(t,x)\rangle^{(0)}=%
\mathcal{O}(t,\Hat{\psi}^3)$ (see Eq.~\eqref{eq:(3.27)} below).

Now, to find the expansion coefficients $\zeta_i^{(1)}(t)$, we substitute the
decompositions~\eqref{eq:(3.11)}--\eqref{eq:(3.13)} and~Eq.~\eqref{eq:(3.8)}
into~Eq.~\eqref{eq:(3.7)} and take the expectation value of both sides
\emph{in the presence of the background fields}. Since the flow-time evolution
of the background fermion field is given by
\begin{align}
   \Hat{\chi}(t,x)=e^{t\Hat{\mathcal{D}}^2}\Hat{\psi}(x)=\Hat{\psi}(x)
   +\mathcal{O}(t),
\label{eq:(3.21)}
\end{align}
we can set $\Hat{\chi}(t,x)\to\Hat{\psi}(x)$ in our calculation which neglects
$\mathcal{O}(t)$ terms. This yields
\begin{align}
   &\left\langle
   \Hat{\psi}^a(x)\left[
   \Hat{\mathcal{D}}_\mu^{ab}b_\nu^b(t,x)-\Hat{\mathcal{D}}_\nu^{ab}b_\mu^b(t,x)
   \right]\right\rangle^{(1)}
   -\left\langle
   \Hat{\psi}^a(x)\left[
   \Hat{\mathcal{D}}_\mu^{ab}a_\nu^b(x)-\Hat{\mathcal{D}}_\nu^{ab}a_\mu^b(x)
   \right]\right\rangle^{(1)}
\notag\\
   &\qquad{}
   +\left\langle
   \Hat{\psi}^a(x)f^{abc}b_\mu^b(t,x)b_\nu^c(t,x)
   -\Hat{\psi}^a(x)f^{abc}a_\mu^b(x)a_\nu^c(x)
   \right\rangle^{(1)}
\notag\\
   &\qquad\qquad{}
   +\left\langle
   k^a(t,x)\Hat{F}_{\mu\nu}^a(x)
   \right\rangle^{(1)}
   -\left\langle
   p^a(x)\Hat{F}_{\mu\nu}^a(x)
   \right\rangle^{(1)}
\notag\\
   &\qquad{}
   +\left\langle
   k^a(t,x)\left[
   \Hat{\mathcal{D}}_\mu^{ab}b_\nu^b(t,x)-\Hat{\mathcal{D}}_\nu^{ab}b_\mu^b(t,x)
   \right]
   -p^a(x)\left[
   \Hat{\mathcal{D}}_\mu^{ab}a_\nu^b(x)-\Hat{\mathcal{D}}_\nu^{ab}a_\mu^b(x)
   \right]\right\rangle^{(1)}
\notag\\
   &=\zeta_1^{(1)}(t)\Hat{\psi}^a(x)\Hat{F}_{\mu\nu}^a(x)
\notag\\
   &\qquad{}
   +\zeta_2^{(1)}(t)
   \left[
   \gamma_\mu\gamma_\rho\Hat{\psi}^a(x)\Hat{F}_{\rho\nu}^a(x)
   -\gamma_\nu\gamma_\rho\Hat{\psi}^a(x)\Hat{F}_{\rho\mu}^a(x)
   \right]
\notag\\
   &\qquad\qquad{}
   +\zeta_3^{(1)}(t)
   \sigma_{\rho\sigma}\sigma_{\mu\nu}\Hat{\psi}^a(x)
   \Hat{F}_{\rho\sigma}^a(x)+\mathcal{O}(t),
\label{eq:(3.22)}
\end{align}
where the superscript~$^{(1)}$ implies that the expectation values are computed
in the one-loop order.

Now, in~Eq.~\eqref{eq:(3.22)}, $\langle b_\nu^b(t,x)\rangle^{(1)}$ in the first
term is a dimension~$1$ combination of background fields that behaves as the
adjoint representation under the background gauge transformation; it possesses
one vector index. The lowest-dimensional operator of such properties
is~$t\Hat{\mathcal{D}}_\mu^{ba}\Hat{F}_{\mu\nu}^a(x)$, which is
already~$\mathcal{O}(t)$; thus we can neglect this term because we are
neglecting $\mathcal{O}(t)$~terms in the present calculation.

On the other hand, by carrying out a calculation of the type explained below,
one finds that $\langle a_\nu^b(x)\rangle^{(1)}$ in the second term
of~Eq.~\eqref{eq:(3.22)} identically vanishes under dimensional regularization,
because there is no mass scale (such as~$t$) that makes the loop integral
nonzero for any~$D$. For the same reason,
$\langle p^a(x)\rangle^{(1)}$ in the third line also identically vanishes under
dimensional regularization.\footnote{More specifically, one ends up with
one-loop integrals of the form~$\int\frac{d^Dp}{(2\pi)^D}\frac{1}{(p^2)^\alpha}$
that identically vanish under dimensional regularization.}

The second line of~Eq.~\eqref{eq:(3.22)} is evaluated as follows: The
tree-level $bb$ propagator \emph{in the presence of the background fields\/} is
given by (Ref.~\cite{Suzuki:2015bqa})
\begin{equation}
   \left\langle b_\mu^a(t,x)b_\nu^b(s,y)\right\rangle_0
   =g_0^2\int_{t+s}^\infty d\xi\,
   \left(e^{\xi\Hat{\Delta}_x}\right)_{\mu\nu}^{ab}\delta(x-y)
   +\mathcal{O}(\Hat{\psi}^2),
\label{eq:(3.23)}
\end{equation}
where
\begin{equation}
   \Hat{\Delta}_{\mu\nu}^{ab}
   =\delta_{\mu\nu}(\Hat{\mathcal{D}}^2)^{ab}
   +2\Hat{\mathcal{F}}_{\mu\nu}^{ab},\qquad
   \Hat{\mathcal{F}}_{\mu\nu}^{ab}
   \equiv f^{acb}\Hat{F}_{\mu\nu}^c(x).
\label{eq:(3.24)}
\end{equation}
In Eq.~\eqref{eq:(3.23)}, we have discarded the contribution coming
from~$\xi=\infty$ because $\Hat{\Delta}_x<0$ at least in perturbation theory.
The tree-level $aa$ propagator is given simply by setting $t=s=0$ in this
expression. Then we have
\begin{align}
   &\left\langle
   \Hat{\psi}^a(x)f^{abc}b_\mu^b(t,x)b_\nu^c(t,x)
   -\Hat{\psi}^a(x)f^{abc}a_\mu^b(x)a_\nu^c(x)
   \right\rangle^{(1)}
\notag\\
   &=\frac{g_0^2}{(4\pi)^2}C_2(G)\frac{-4}{D-4}(8\pi t)^{2-D/2}
   \Hat{\psi}^a(x)\Hat{F}_{\mu\nu}^a(x).
\label{eq:(3.25)}
\end{align}
To obtain this, we first use the above $bb$ and $aa$ propagators for the
contractions in~Eq.~\eqref{eq:(3.25)}. Then we express the delta function
in~Eq.~\eqref{eq:(3.23)} as $\delta(x-y)=\int\frac{d^Dp}{(2\pi)^D}\,e^{ip(x-y)}$
and shift the plain wave~$e^{ipx}$ in the left of the differential
operators (Ref.~\cite{Suzuki:2015bqa}). Finally the loop momentum integration
yields Eq.~\eqref{eq:(3.25)}.

The fourth term of~Eq.~\eqref{eq:(3.22)} is evaluated as follows: We see that
the tree-level $kb$ propagator is given by
\begin{align}
   &\left\langle k^a(t,x)b_\mu^b(s,y)\right\rangle_0
\notag\\
   &=-g_0^2
   \left(e^{t\Hat{\Slash{\mathcal{D}}}_x^2}
   \frac{1}{\Hat{\Slash{\mathcal{D}}}_x}\right)^{ac}f^{cde}\gamma_\nu
   \Hat{\psi}^e(x)
   \int_s^\infty du\,
   \left(e^{u\Hat{\Delta}_x}\right)_{\nu\mu}^{db}\delta(x-y)
   +\mathcal{O}(\Hat{D}\Hat{\psi},\Hat{\psi}^3).
\label{eq:(3.26)}
\end{align}
Then using Eqs.~\eqref{eq:(3.23)} and~\eqref{eq:(3.26)} for the contractions in
the solution of the flow equation (Ref.~\cite{Suzuki:2015bqa}),
\begin{align}
   k^a(t,x)
   &=\left(e^{t\Hat{\mathcal{D}}^2}\right)^{ab}p^b(x)
\notag\\
   &\qquad{}
   +\int_0^tds\,
   \left[e^{(t-s)\Hat{\mathcal{D}}^2}\right]^{ab}
   \left[
   2f^{bcd}b_\mu^c(s,x)\Hat{\mathcal{D}}_\mu^{de}
   +f^{bcd}f^{dfe}b_\mu^c(s,x)b_\mu^f(s,x)
   \right]
\notag\\
   &\qquad\qquad\qquad\qquad\qquad\qquad{}
   \times\left\{
   \left[e^{s\Hat{\mathcal{D}}^2}\right]^{eg}\Hat{\psi}^g(x)+k^e(s,x)\right\},
\label{eq:(3.27)}
\end{align}
we have
\begin{equation}
   \left\langle
   k^a(t,x)\Hat{F}_{\mu\nu}^a(x)
   \right\rangle^{(1)}
   =\frac{g_0^2}{(4\pi)^2}C_2(G)\frac{2}{D-4}(8\pi t)^{2-D/2}
   \Hat{\psi}^a(x)\Hat{F}_{\mu\nu}^a(x).
\label{eq:(3.28)}
\end{equation}

Finally, after some careful calculation using the above relations, we have
\begin{align}
   &\left\langle
   k^a(t,x)\left[
   \Hat{\mathcal{D}}_\mu^{ab}b_\nu^b(t,x)-\Hat{\mathcal{D}}_\nu^{ab}b_\mu^b(t,x)
   \right]
   -p^a(x)\left[
   \Hat{\mathcal{D}}_\mu^{ab}a_\nu^b(x)-\Hat{\mathcal{D}}_\nu^{ab}a_\mu^b(x)
   \right]\right\rangle^{(1)}
\notag\\
   &=\frac{g_0^2}{(4\pi)^2}C_2(G)\frac{2}{(D-4)(D-2)D}(8\pi t)^{2-D/2}
\notag\\
   &\qquad{}
   \times
   \left[
   D\gamma_\mu\gamma_\rho\Hat{\psi}^a(x)\Hat{F}_{\rho\nu}^a(x)
   -D\gamma_\nu\gamma_\rho\Hat{\psi}^a(x)\Hat{F}_{\rho\mu}^a(x)
   +2\sigma_{\rho\sigma}\sigma_{\mu\nu}\Hat{\psi}^a(x)\Hat{F}_{\rho\sigma}^a(x)
   \right].
\label{eq:(3.29)}
\end{align}

Thus, from~Eqs.~\eqref{eq:(3.22)}, \eqref{eq:(3.25)}, \eqref{eq:(3.28)},
and~\eqref{eq:(3.29)}, we have the one-loop coefficients,
\begin{align}
   \zeta_1^{(1)}(t)&=\frac{g_0^2}{(4\pi)^2}C_2(G)\frac{-2}{D-4}(8\pi t)^{2-D/2},
\label{eq:(3.30)}
\\
   \zeta_2^{(1)}(t)
   &=\frac{g_0^2}{(4\pi)^2}C_2(G)\frac{2}{(D-4)(D-2)}(8\pi t)^{2-D/2},
\label{eq:(3.31)}
\\
   \zeta_3^{(1)}(t)&=\frac{g_0^2}{(4\pi)^2}C_2(G)
   \frac{4}{(D-4)(D-2)D}(8\pi t)^{2-D/2}.
\label{eq:(3.32)}
\end{align}
Note that these coefficients themselves have a pole at~$D=4$.

\subsection{Final steps}
Substituting Eqs.~\eqref{eq:(3.30)}--\eqref{eq:(3.32)}
into~Eq.~\eqref{eq:(3.10)}, we have an expression for the supercurrent:
\begin{align}
   S_{\mu R}(x)
   &=-\frac{1}{2g_0}
   \left[1+\frac{g_0^2}{(4\pi)^2}C_2(G)\frac{2(D-18)}{(D-2)D}(8\pi t)^{2-D/2}
   \right]
   \sigma_{\rho\sigma}\gamma_\mu\chi^a(t,x)G_{\rho\sigma}^a(t,x)
\notag\\
   &\qquad{}
   +\frac{1}{2g_0}
   \frac{g_0^2}{(4\pi)^2}C_2(G)\frac{8(D-10)}{(D-2)D}(8\pi t)^{2-D/2}
   \gamma_\nu\chi^a(t,x)G_{\nu\mu}^a(t,x)+\mathcal{O}(t)+\mathcal{O}(g_0^3).
\label{eq:(3.33)}
\end{align}
We may rewrite this in terms of renormalized quantities. The renormalized gauge
coupling in the MS scheme is given by
\begin{equation}
   g_0^2=\mu^{2\epsilon}g^2
   \left[1+\frac{g^2}{(4\pi)^2}C_2(G)\frac{1}{\epsilon}(-3)+\mathcal{O}(g^4)
   \right].
\label{eq:(3.34)}
\end{equation}
For the gaugino field, we use a ``ringed variable''
(Ref.~\cite{Makino:2014taa}):
\begin{align}
   \mathring{\chi}(t,x)
   &=\sqrt{\frac{-\dim(G)}
   {(4\pi)^2t^2
   \left\langle\Bar{\chi}(t,x)
   \overleftrightarrow{\Slash{\mathcal{D}}}\chi(t,x)
   \right\rangle}}
   \,\chi(t,x)
\notag\\
   &=\frac{1}{(8\pi t)^{\epsilon/2}}
   \left\{1+\frac{g^2}{(4\pi)^2}C_2(G)
   \left[\frac{3}{2}\frac{1}{\epsilon}+\frac{3}{2}\ln(8\pi\mu^2 t)
   -\frac{1}{2}\ln(432)\right]
   +\mathcal{O}(g^4)\right\}
   \chi(t,x),
\label{eq:(3.35)}
\end{align}
where
\begin{equation}
   \overleftrightarrow{\mathcal{D}}_\mu
   \equiv\mathcal{D}_\mu-\overleftarrow{\mathcal{D}}_\mu,
\label{eq:(3.36)}
\end{equation}
which is free from the wave-function renormalization of the flowed fermion
field (see Ref.~\cite{Luscher:2013cpa}). Then, we have
\begin{align}
   S_{\mu R}(x)
   &=-\frac{1}{2g}
   \left\{1+\frac{g^2}{(4\pi)^2}C_2(G)
   \left[-\frac{7}{2}-\frac{3}{2}\ln(8\pi\mu^2 t)+\frac{1}{2}\ln(432)\right]
   \right\}
   \sigma_{\rho\sigma}\gamma_\mu\mathring{\chi}^a(t,x)G_{\rho\sigma}^a(t,x)
\notag\\
   &\qquad{}
   -\frac{g}{(4\pi)^2}C_2(G)3
   \gamma_\nu\mathring{\chi}^a(t,x)G_{\nu\mu}^a(t,x)+\mathcal{O}(t)
   +\mathcal{O}(g^3).
\label{eq:(3.37)}
\end{align}
This expression is manifestly UV finite because local products of the flowed
gauge field and the ringed fermion field are free from UV divergences. This
must be so because the supercurrent is a physical Noether current that must be
free from UV divergences.

Finally, since $S_{\mu R}(x)$ is totally composed from bare quantities
as, e.g., Eq.~\eqref{eq:(3.33)} shows, it is independent of the
renormalization scale~$\mu$ when expressed by the running gauge
coupling~$\Bar{g}(\mu)$, defined by
\begin{equation}
   \mu\frac{d\Bar{g}(\mu)}{d\mu}=\beta(\Bar{g}(\mu)),\qquad
   \beta(g)\equiv\lim_{\epsilon\to0}
   \left.\mu\frac{\partial}{\partial\mu}g\right|_{\text{$g_0$ fixed}},
\label{eq:(3.38)}
\end{equation}
explicitly,
\begin{equation}
   \beta(g)=-b_0g^3-b_1g^5+\mathcal{O}(g^7),\qquad
   b_0=\frac{1}{(4\pi)^2}3C_2(G),\qquad
   b_1=\frac{1}{(4\pi)^4}6C_2(G)^2.
\label{eq:(3.39)}
\end{equation}
Thus we set~$\mu=1/\sqrt{8t}$. Then,
\begin{align}
   S_{\mu R}(x)
   &=-\frac{1}{2\Bar{g}(1/\sqrt{8t})}
   \left\{1+\frac{\Bar{g}(1/\sqrt{8t})^2}{(4\pi)^2}C_2(G)
   \left[-\frac{7}{2}-\frac{3}{2}\ln\pi+\frac{1}{2}\ln(432)\right]
   \right\}
\notag\\
   &\qquad\qquad\qquad\qquad\qquad\qquad\qquad\qquad
   \qquad\qquad\qquad{}\times
   \sigma_{\rho\sigma}\gamma_\mu\mathring{\chi}^a(t,x)G_{\rho\sigma}^a(t,x)
\notag\\
   &\qquad{}
   -\frac{\Bar{g}(1/\sqrt{8t})}{(4\pi)^2}C_2(G)3
   \gamma_\nu\mathring{\chi}^a(t,x)G_{\nu\mu}^a(t,x)+\mathcal{O}(t)
   +\mathcal{O}(\Bar{g}(1/\sqrt{8t})^3).
\label{eq:(3.40)}
\end{align}
This shows that the perturbative determination of the expansion coefficients
is justified for~$t\to0$. Taking the $t\to0$ limit to get rid of higher-order
terms, we arrive at the announced expression,
\begin{align}
   S_{\mu R}(x)
   &=\lim_{t\to0}\biggl(
   -\frac{1}{2\Bar{g}(1/\sqrt{8t})}
   \left\{1+\frac{\Bar{g}(1/\sqrt{8t})^2}{(4\pi)^2}C_2(G)
   \left[-\frac{7}{2}-\frac{3}{2}\ln\pi+\frac{1}{2}\ln(432)\right]
   \right\}
\notag\\
   &\qquad\qquad\qquad\qquad\qquad\qquad\qquad\qquad\qquad\qquad\qquad{}
   \times\sigma_{\rho\sigma}\gamma_\mu\mathring{\chi}^a(t,x)G_{\rho\sigma}^a(t,x)
\notag\\
   &\qquad\qquad{}
   -\frac{\Bar{g}(1/\sqrt{8t})}{(4\pi)^2}C_2(G)3
   \gamma_\nu\mathring{\chi}^a(t,x)G_{\nu\mu}^a(t,x)\biggr).
\label{eq:(3.41)}
\end{align}
If one prefers the $\overline{\text{MS}}$ scheme instead of the MS scheme
assumed in this expression, it suffices to make the replacement
\begin{equation}
   \ln\pi\to\gamma-2\ln2,
\label{eq:(3.42)}
\end{equation}
where $\gamma$ is Euler's constant.\footnote{The factor $\ln\pi$
in~\eqref{eq:(3.41)} comes from the product of the factor~$(4\pi)^\epsilon$
arising from the one-loop momentum integral and the pole $1/\epsilon$ as
$1/\epsilon+\ln\pi+\ln4$. Since the gauge couplings in the MS scheme and in
the $\overline{\text{MS}}$ scheme are related as
\begin{equation}
   g_{\text{MS}}^2=\pi^{-\epsilon}e^{\epsilon(\gamma-2\ln2)}g_{\overline{\text{MS}}}^2,
\end{equation}
the factor~$(4\pi)^\epsilon$ is replaced by~$(4e^{\gamma-2\ln2})^\epsilon$ in
the $\overline{\text{MS}}$ scheme, thus resulting in the
rule~\eqref{eq:(3.42)}.}


\section{Conclusion}
\label{sec:4}
In this paper, we have obtained a representation of the properly normalized
conserved supercurrent in the 4D $\mathcal{N}=1$ SYM, in terms of the small
flow-time expansion of the gradient flow, Eq.~\eqref{eq:(3.41)}. Because of
remarkable renormalization properties of the gradient flow
(Refs.~\cite{Luscher:2011bx,Luscher:2013cpa,Hieda:2016xpq}), this
representation possesses a meaning independent of the adopted regularization.
This in particular implies that the representation can be used in lattice
numerical simulations, as a similar representation of the energy--momentum
tensor can be
(Refs.~\cite{Asakawa:2013laa,Taniguchi:2016ofw,Kitazawa:2016dsl}). An important
application would be to determine the supersymmetric point in the parameter
space, for which the conservation of the supercurrent provides a definite
criterion.

For more general supersymmetric theories, one has to take into account the
flow of the scalar field. We may adopt a simple flow equation,
\begin{equation}
   \partial_t\varphi(t,x)=D_\mu D_\mu\varphi(t,x),\qquad\varphi(t=0,x)=\phi(x),
\label{eq:(4.1)}
\end{equation}
because the inclusion of further terms corresponding to mass, self-interaction,
and Yukawa terms in the right-hand side would break the renormalizability
(Ref.~\cite{Capponi:2015ucc}). By using this setup, it must be possible to
obtain a representation of the properly normalized conserved supercurrent in
general supersymmetric theories; for general theories the parameter tuning in
lattice numerical simulations will be really demanding. We hope to study this
problem in the near future.

\section*{Acknowledgements}
We would to thank David B. Kaplan and Tetsuya Onogi for discussions which
motivated the present work.
This work was supported by JSPS KAKENHI Grant Numbers 16J02259 (A.~K.)
and~16H03982 (H.~S.).

\section*{Acknowledgements}
We would like to thank David B. Kaplan and Tetsuya Onogi for discussions that
motivated the present work.
This work was supported by JSPS KAKENHI Grant Numbers 16J02259 (A.~K.)
and~16H03982 (H.~S.).

\appendix

\section{Charge conjugation matrix in dimensional regularization}
\label{app:A}
In this appendix, we consider the charge conjugation matrix in dimensional
regularization. In the $D=4$ Euclidean space, the charge conjugation matrix~$C$
is defined such that
\begin{equation}
   C^{-1}\gamma_\mu C=-\gamma_\mu^T,\qquad C^T=-C,
\label{eq:(A1)}
\end{equation}
where $T$ denotes the transpose and thus
\begin{equation}
   C^{-1}\sigma_{\mu\nu}C=-\sigma_{\mu\nu}^T,\qquad C^{-1}\gamma_5C=\gamma_5^T.
\label{eq:(A2)}
\end{equation}
For a general complex spacetime dimension~$D$, we postulate
\begin{equation}
   C^{-1}\gamma_\mu C=-s_1(D)\gamma_\mu^T,\qquad C^T=-s_2(D)C,
\end{equation}
where coefficients~$s_1(D)$ and~$s_2(D)$ are meromorphic functions of~$D$.
Requiring the usual properties of the transpose, we find
\begin{equation}
   s_1(D)^2=s_2(D)^2=1.
\end{equation}
However, since $s_1(D=4)=s_2(D=4)=+1$, we have $s_1(D)=s_2(D)=+1$ for
general~$D$; the charge conjugation matrix in dimensional regularization also
satisfies Eqs.~\eqref{eq:(A1)} and~\eqref{eq:(A2)}. These relations are fully
employed in our computation in the main text.

It should be noted that the above definition obtained by the analytic
continuation from~$D=4$ does \emph{not\/} necessarily coincide with the
conventional charge conjugation matrix at a fixed integer dimension. For
example, for~$D=5$, we have $C^{-1}\gamma_\mu C=+\gamma_\mu^T$ as implied by the
latter relation of~Eq.~\eqref{eq:(A2)}. Nevertheless, the above definition is
perfectly legitimate from the perspective of dimensional regularization.

\section{Two-loop order improvement through the superconformal anomaly}
\label{app:B}
In Ref.~\cite{Suzuki:2013gza} on the energy--momentum tensor in pure
Yang--Mills theory, it was possible to improve the small flow-time
representation by using the information of the trace (or conformal) anomaly to
the two-loop order. We can imitate this strategy for the supercurrent in the
present 4D $\mathcal{N}=1$ SYM as follows.

We thus require that a formula such as~Eq.~\eqref{eq:(3.37)} reproduces the
superconformal anomaly (Refs.~\cite{Abbott:1977in,Curtright:1977cg,%
Inagaki:1978iu,Majumdar:1980ej,Nicolai:1980km,Hagiwara:1979pu,Hagiwara:1980ys,%
Kumar:1982ng,Nakayama:1983qt}),
\begin{equation}
   \gamma_\mu S_{\mu R}(x)
   =-\frac{\beta(g)}{g^2}\{\sigma_{\mu\nu}\psi^a F_{\mu\nu}^a\}_R(x),
\label{eq:(B1)}
\end{equation}
to the two-loop order, where the beta function~$\beta(g)$ is given
by~Eq.~\eqref{eq:(3.39)}. First, we need to know the expression for the
renormalized composite operator in the right-hand side,
$\{\sigma_{\mu\nu}\psi^a F_{\mu\nu}^a\}_R(x)$, e.g., in the MS scheme. For
symmetry reasons, this operator is multiplicatively renormalized. To find the
renormalization constant, we note the relation
\begin{align}
   &\left(1-\frac{3}{2}\Delta\right)\sigma_{\mu\nu}\psi^a(x)F_{\mu\nu}^a(x)
\notag\\
   &=\left\{1+\frac{g^2}{(4\pi)^2}C_2(G)
   \left[2+\frac{3}{2}\ln(8\pi\mu^2 t)+\frac{1}{2}\ln(432)\right]
   \right\}\sigma_{\mu\nu}
   \mathring{\chi}^a(t,x)G_{\mu\nu}^a(t,x)+\mathcal{O}(t),
\label{eq:(B2)}
\end{align}
where $\Delta$ is defined in~Eq.~\eqref{eq:(2.27)}, which follows
from~Eqs.~\eqref{eq:(3.9)}, \eqref{eq:(3.30)}--\eqref{eq:(3.32)},
and~\eqref{eq:(3.35)} to the one-loop order. Since the right-hand side of this
relation is manifestly finite, this is the renormalized composite
operator~$\{\sigma_{\mu\nu}\psi^a F_{\mu\nu}^a\}_R(x)$ in the MS scheme. Having
obtained this information, we see that the expression
\begin{align}
   S_{\mu R}(x)
   &=-\frac{1}{2g}
   \left\{1+\frac{g^2}{(4\pi)^2}C_2(G)
   \left[-\frac{7}{2}-\frac{3}{2}\ln(8\pi\mu^2 t)+\frac{1}{2}\ln(432)\right]
   \right\}
   \sigma_{\rho\sigma}\gamma_\mu\mathring{\chi}^a(t,x)G_{\rho\sigma}^a(t,x)
\notag\\
   &\qquad{}
   -b_0g\biggl\{
   1+\frac{b_1}{b_0}g^2
   +\frac{g^2}{(4\pi)^2}C_2(G)
   \left[2+\frac{3}{2}\ln(8\pi\mu^2 t)+\frac{1}{2}\ln(432)\right]
   +\mathcal{O}(g^4)\biggr\}
\notag\\
   &\qquad\qquad\qquad{}
   \times\gamma_\nu\mathring{\chi}^a(t,x)G_{\nu\mu}^a(t,x)+\mathcal{O}(t),
\label{eq:(B3)}
\end{align}
reproduces Eq.~\eqref{eq:(B1)} to the two-loop order. Finally, repeating the
renormalization group argument that lead to~Eq.~\eqref{eq:(3.41)}, we have
\begin{align}
   S_{\mu R}(x)
   &=\lim_{t\to0}\biggl(
   -\frac{1}{2\Bar{g}(1/\sqrt{8t})}
   \left\{1+\frac{\Bar{g}(1/\sqrt{8t})^2}{(4\pi)^2}C_2(G)
   \left[-\frac{7}{2}-\frac{3}{2}\ln\pi+\frac{1}{2}\ln(432)\right]
   \right\}
\notag\\
   &\qquad\qquad\qquad\qquad\qquad\qquad\qquad\qquad\qquad{}
   \times\sigma_{\rho\sigma}\gamma_\mu\mathring{\chi}^a(t,x)G_{\rho\sigma}^a(t,x)
\notag\\
   &\qquad\qquad{}
   -\frac{\Bar{g}(1/\sqrt{8t})}{(4\pi)^2}C_2(G)3\left\{
   1+\frac{\Bar{g}(1/\sqrt{8t})^2}{(4\pi)^2}C_2(G)
   \left[4+\frac{3}{2}\ln\pi+\frac{1}{2}\ln(432)\right]\right\}
\notag\\
   &\qquad\qquad\qquad\qquad\qquad\qquad\qquad\qquad\qquad\qquad{}
   \times\gamma_\nu\mathring{\chi}^a(t,x)G_{\nu\mu}^a(t,x)\biggr).
\label{eq:(B3)}
\end{align}

\end{document}